%% file: ClusterLens.tex
\documentclass[prl,aps,twocolumn,superscriptaddress,preprintnumbers,nofootinbib]{revtex4}
\pdfoutput=1

\usepackage[pdftex]{graphicx}
\usepackage[latin1]{inputenc}
\usepackage{amsmath}
\usepackage{amsfonts}
\usepackage{amssymb}
\usepackage{rotating}
\usepackage{subfigure}
\usepackage{paralist} 
\usepackage{verbatim}
\usepackage{color}
\usepackage{soul} 
\usepackage{appendix}

\usepackage{natbib}
\usepackage{epsfig,hyperref}
\usepackage{array}
\usepackage{threeparttable}

\newcommand{\PredictedSigAllNullTen}{4.2}
\newcommand{\SigAllTheoryFour}{2.9}
\newcommand{\SigAllTheoryTen}{3.2}
\newcommand{\SigAllNullFour}{3.6}
\newcommand{\SigAllNullTen}{3.8}
\newcommand{\SigDeepOneFour}{2.0}
\newcommand{\SigDeepFiveFour}{3.6}
\newcommand{\SigDeepSixFour}{1.3}

\newcommand{\bestfitmass}{$M_{200\bar{\rho}_0}=(2.0\pm 0.7) \times 10^{13}~h^{-1}M_\odot$}
\newcommand{\bestfitconcentration}{$c_{200\bar{\rho}}=(5.4\pm0.8)$}
\newcommand{\bestfitgoodness}{$\chi^2/\nu=1.5$}

\newcommand*\Bell{\ensuremath{\boldsymbol l}}

\usepackage{color}
\definecolor{orange}{rgb}{1,0.3,0}

\begin{document}

\title{Evidence of Lensing of the Cosmic Microwave Background by Dark Matter Halos}
\date{\today}

%\author{ACTPol Collaboration $^{1}$}
%\author{ACTPol Collaboration}
\input{authors.tex}

%\altaffiltext{1}{ACT Institutes} ---------
\begin{abstract}
We present evidence of the gravitational lensing of the
cosmic microwave background by $10^{13}$ solar mass dark matter halos.
Lensing convergence maps from the Atacama Cosmology Telescope Polarimeter
(ACTPol) are stacked at the positions of around 12,000 optically-selected
CMASS galaxies from the SDSS-III/BOSS survey. The mean lensing signal is consistent with
simulated dark matter halo profiles, and is favored over a null signal
at $3.2\sigma$ significance. This result demonstrates the potential of
microwave background lensing to probe the dark matter distribution in
galaxy group and galaxy cluster halos.
\end{abstract}

\maketitle

\section{Introduction}
\label{sec:intro}
\setcounter{footnote}{0} % must be after first \section!

Measuring the gravitational lensing of the cosmic microwave background (CMB) by intervening structure is a potentially powerful way to map out the mass distribution in the Universe.  Advantages of CMB lensing over lensing measured at other wavelengths include that the CMB is a source that fills the whole sky, is at a known redshift, and has well understood statistical properties.
To date, the lensing of the CMB caused by the large-scale projected dark matter distribution has been observed by a number of CMB experiments with ever increasing statistical significance \cite{smith/etal/2007, hirata/etal/2008, das/etal/2011, vanengelen/etal/2012,planck_lens/2013}.  This lensing signal has been detected in both CMB temperature and polarization maps and in cross-correlation with other tracers of large-scale structure \cite{smith/etal/2007, hirata/etal/2008, sherwin/etal/2012, bleem/etal/2012, geach/etal/2013, planck_lens/2013, holder/etal/2013, planck_ciblensing/2013, hanson/etal/2013, pbear-herschel/2013, pbear-eeeb/2013, hand/etal/2013, bianchini/etal/2014, dipompeo/etal/2014, fornego/etal/2014,HillSpergelLensing2014}.  These CMB lensing measurements have become precise enough that they now provide interesting constraints on a number of cosmological parameters such as curvature and the amplitude of matter fluctuations \cite{planck_params/2013}.    These constraints can be expected to significantly improve with the advent of near-term and next-generation CMB datasets \cite{CMBS4Nus,SnowmassInf2013,calabrese2014precision}.

Previous studies have focused on the lensing of the CMB by large-scale structure corresponding to scales between tens and several hundred comoving Mpc. As the data improve it is possible to shift focus to smaller scales, particularly those which have undergone appreciable nonlinear growth.  On small enough scales, the CMB is lensed by individual dark matter halos.
We refer to this small-scale signal as ``CMB halo lensing,'' and note that this lensing can be due to individual galaxy clusters, galaxy groups, and massive galaxies.  Before now, CMB experiments did not have the sensitivity or resolution to detect this signal which was hypothesized to exist over a decade ago~\cite{seljak2000lensing,zaldarriaga2000lensing,dodelsonGalaxy2003,holder2004gravitational,dodelson2004cmb,ValeAmblardWhite2004,maturi.bartelmann.ea:2005,LewisKing2006,HuHolzVale2007,hu2007cluster,Yoo2008,Yoo2010,Melin2014}.

In this work, we present evidence of the CMB halo lensing signal using the first season of data from ACTPol.  This detection is made by stacking ACTPol reconstructed convergence maps at the positions of CMASS galaxies that have been optically selected from the Sloan Digital Sky Survey-III Baryon Oscillation Spectroscopic Survey Tenth Data Release (SDSS-III/BOSS DR10) (\cite{eisenstein2011sdss,dawson2013baryon,ahn2014tenth}).  This signal is detected at a significance of \SigAllTheoryTen$\sigma$ when we combine the nighttime data from three ACTPol first-season survey regions. We see an excess of \SigDeepSixFour$\sigma$ or greater in each indiviudual survey region, although all fields are needed to give a statistical detection.

\section{CMB Data}
\label{sec:cmbdata}

ACT is located in Parque Astron\'omico Atacama in northern Chile at an altitude of 5190 m.  The 6-meter primary mirror has a resolution of 1.4 arcminutes at a wavelength of 2 millimeters.  Its first polarization-sensitive camera, ACTPol, is described in detail in \cite{niemack/etal/2010} and \cite{Naess2014}.  ACTPol observed from Sept. 11 to Dec. 14, 2013 at 146 GHz. 
Four ``deep field'' patches were surveyed near the celestial equator at right ascensions of $150^\circ$, $175^\circ$, $355^\circ$, and $35^\circ$, which we call D1 (73 deg$^2$), D2 (70 deg$^2$), D5 (70 deg$^2$), and D6 (63 deg$^2$).  The scan strategy allows for each patch to be observed in a range of different parallactic angles while scanning horizontally, which aids in separating instrumental effects from celestial polarization.  White noise map sensitivity levels for the patches are 16.2, 17, 13.2, and 11.2 $\mu$K-arcmin respectively in temperature, with polarization noise levels higher by roughly $\sqrt{2}$.  All patches were observed during nighttime hours for some fraction of the time.  The nighttime data fraction is $50\%, 25\%, 76\%$, and $94\%$ for D1, D2, D5, and D6 respectively.  We use only nighttime data from D1, D5, and D6 in this analysis.  Further details about the observations and mapmaking can be found in \cite{Naess2014}.

We template-subtract point sources from these maps by filtering the D1, D5, and D6 patches with a filter matched to the ACTPol beam profile. Point sources with a signal at least five times larger than the background uncertainty in the filtered maps are identified, and their fluxes are measured.  A template of beam-convolved point sources is then constructed for each patch and subsequently subtracted from the corresponding patch.  As a result, point sources with fluxes above 8 mJy are removed from D1, and sources with fluxes above 5 mJy are removed from D5 and D6.

Overall calibration of the ACTPol patches is achieved by comparing to the Planck 143 GHz temperature map \cite{planck_mission/2013} and following the method described in \cite{louis/etal/2014}.  The patches are then multiplied by a factor of 1.012 to correspond to the WMAP calibration as in \cite{Naess2014}.

\section{Optical Data}
\label{sec:opticaldata}

SDSS I and II  obtained imaging data of 11,000 deg$^2$ using the 2.5-meter SDSS Telescope \cite{york2000sloan,gunn20062}.  This survey has five photometric bands.  SDSS-III BOSS extended this imaging survey by 3,000 deg$^2$ \cite{eisenstein2011sdss}.  Based on the resulting photometric catalog of galaxies, CMASS (``constant mass'') galaxies were selected extending the luminous red galaxy (LRG) selection of \cite{Eisenstein2001} to bluer and fainter galaxies. These galaxies form a roughly volume-limited sample with $z>0.4$ and satisfy the criterion that their number density be high enough to probe large-scale structure at redshifts of about 0.5 \cite{andersonCMASS2012}.  The BOSS spectroscopic survey targeted these galaxies obtaining spectroscopic redshifts, and these galaxies have been used in a number of cosmological analyses \cite{andersonCMASS2012, reidCMASS2012}.

Using the tenth SDSS public data release (DR10), we selected CMASS galaxies from the BOSS catalog.\footnote{\url{https://data.sdss3.org/datamodel/files/SPECTRO_REDUX/specObj.html}. We used the keywords \texttt{BOSS\_TARGET1 \&\& 2, SPECPRIMARY == 1, ZWARNING\_NOQSO == 0}, and \texttt{(CHUNK != "boss1") \&\& (CHUNK != "boss2")}. The keywords are described here: \url{https://www.sdss3.org/dr10/algorithms/boss_galaxy_ts.php}}  This selection resulted in 6144,  5211, and 5420 CMASS galaxies that lie within D1, D5, and D6 respectively.    These galaxies span a redshift range of about $z=0.4$ to $z=0.7$, with a mean redshift of $z=0.54$.  The galaxies were cross referenced with galaxies in the SDSS-III photometric catalog,{\footnote{\url{http://data.sdss3.org/datamodel/files/BOSS_PHOTOOBJ/RERUN/RUN/CAMCOL/photoObj.html}}} using a shared galaxy identification number, to obtain more accurate celestial position information.  

A subset of CMASS galaxies have optical weak-lensing mass estimates of their average halo masses using the publicly-available CFHTLenS galaxy catalog \cite{miyatake2013weak,cfhtlens2012}.  This subset has an additional redshift cut of $z \in [0.47, 0.59]$ and a stellar mass cut of $10^{11.1}~h_{70}^{-2} M_{\odot} < M_{\star} < 10^{12.0}~h_{70}^{-2} M_{\odot}$ relative to the full CMASS sample.\footnote{The full CMASS sample has a stellar mass range of roughly $10^{10.6}~h_{70}^{-2} M_{\odot} < M_{\star} < 10^{12.2}~h_{70}^{-2} M_{\odot}$.}  The average halo mass estimate for this CMASS galaxy subsample is $M_{200\bar{\rho}_0} = (2.3 \pm 0.1) \times 10^{13}~h^{-1} M_{\odot}$  \cite{miyatake2013weak}, where $M_{200\bar{\rho}_0}$ is defined as the mass within $R_{200}$, a radius within which the average density is 200 times the mean density of matter today.  If we had adopted the additional redshift and stellar mass cuts of this subsample of CMASS galaxies, then the number of galaxies falling in the ACTPol patches would have been reduced by roughly a factor of two;  so we instead stack on the full CMASS galaxy sample within our survey regions for this work.  

Since we cut out a $70' \times 70'$ `stamp' centered on each CMASS galaxy from the ACTPol temperature maps, we exclude all galaxies whose stamp does not fall entirely within the corresponding ACTPol patch. We find from simulations that this stamp size is roughly the minimum required to obtain unbiased lensing reconstructions using the pipeline described here. We also note that performing reconstructions on small stamps allows us to obtain the necessary precision for the mean field subtraction described in the next section. To avoid noisy parts of the ACTPol patches, we also remove galaxies for which the mean value of its corresponding inverse variance weight stamp is lower than 0.7, 0.3, and 0.3 times the mean of the weight map of the full patch for D1, D5, and D6 respectively. These factors were chosen so that all of the stamps in our stacks had an average detector hit count above the same minimum value. These cuts leave 4400, 3665, and 4032 galaxies to stack on in D1, D5, and D6 respectively.

\section{Pipeline}
\label{sec:pipeline}
The analysis pipeline used in this work is as follows.  We set the mean of each galaxy-centered $70' \times 70'$ stamp to zero to prevent leakage of power on scales larger than the stamp size due to windowing effects.  Each stamp is then multiplied by an apodization window that consists of the corresponding inverse variance weight stamp that has been smoothed and tapered with a cosine window of width 14 arcminutes.  Each of the stamps is then beam-deconvolved and filtered with the quadratic filter given in \cite{hu2007cluster}.

The filter is constructed by noting that lensing of the CMB temperature field shifts the unlensed temperature field, $\tilde{T}(\hat{\bf n})$, to the lensed temperature field, $T(\hat{{\bf n}})$, so that

\begin{equation}
T(\hat{\bf n}) = \tilde{T}(\hat{{\bf n}} + \nabla{\phi})
\end{equation}
where $\phi$ is the deflection potential and $\nabla \phi$ is the deflection angle. The lensing convergence, $\kappa$, is given by
\begin{equation}
\nabla^2 \phi = -2 \kappa.
\end{equation}
On the arcminute scales of individual dark matter halos, the unlensed CMB can be approximated as a gradient, and lensing induced by the halo alters the CMB field along this gradient direction.  Thus, we search for this signal by looking for deflections correlated with the background CMB gradient. In order to do this, we reconstruct the lensing convergence field, $\kappa$, by constructing two filtered versions of the data: one that is filtered to isolate the background gradient and one that is filtered to isolate small-scale CMB fluctuations. Then we take the divergence of the product of these two maps as described in~\cite{hu2007cluster} and summarized below.  

The first filtered map is constructed by taking the weighted gradient of the lensed CMB map
\begin{equation}
{\bf G}_{\Bell}^{TT} = i \, {\Bell}\, W_l^{TT} \,T_{\Bell},
\end{equation}
where the weight filter is
\begin{equation}
W_l^{TT} = \tilde{C}_l^{TT} (C_l^{TT}+N_l^{TT})^{-1}
\end{equation}
for $l \leq l_{\rm{G}}$, and $W_{l}^{TT} = 0$ for $l > l_{\rm{G}}$. Note that $ \tilde{C}_l$ and ${C}_l$ are the unlensed and lensed CMB power spectra respectively from a fiducial theoretical model based on {\it{Planck}} best-fit parameters, and $N_l$ is the noise power.  Here $l_{\rm{G}}$ is a cutoff scale and is set to $l_{\rm{G}}=2000$. We choose this cutoff since, as shown in ~\cite{hu2007cluster}, the unlensed CMB gradient does not have contributions above $l=2000$, and we want to remove smaller-scale fluctuations. This cutoff in the gradient filter is the main difference between the filter used in this work and the filter used for large-scale structure lensing \cite{hu2002mass}.  When the convergence, $\kappa$, is large (of order 1), as it is for clusters, only the filter with the gradient cutoff returns an unbiased estimate of the convergence \cite{hu2007cluster}.   For smaller convergence values, as measured for galaxy groups in this work, both filters return similar results.

The second filtered map is an inverse-variance weighted map given by
\begin{equation}
L^T_{\Bell} = W_l^T\,T_{\Bell},
\end{equation}
where
\begin{equation}
W_l^T = (C_l^{TT}+N_l^{TT})^{-1}.
\end{equation}

Taking the divergence of the product of these filtered maps, as prescribed in \cite{hu2007cluster}, gives,
\begin{equation}
\frac{\kappa_{\Bell}^{TT}}{A_l^{TT}} = -\int {\rm d}^2 \hat{\bf n}\,e^{-i \hat{\bf n} \cdot {\Bell}} \, \left\{ \nabla \cdot [{\bf G}^{TT}(\hat{\bf n}) \,L^{T}(\hat{\bf n})]\right\}.
\label{eq:main}
\end{equation}
Here the real-space lensing convergence field constructed from temperature data is
\begin{equation}
\kappa^{TT}(\hat{\bf n}) = \int \frac{{\rm d}^2 l}{(2\pi)^2}\,e^{i {\Bell} \cdot \hat{\bf n}}\,\kappa_{\Bell}^{TT}.
\end{equation}
The normalization factor is given by
\begin{equation}
\frac{1}{A_l^{TT}} = \frac{2}{l^2}\, \int \frac{{\rm d}^2 l_1}{(2\pi)^2}\, [{\Bell} \cdot {\Bell}_1] \, W_{l_1}^{TT}\,W_{l_2}^T\, f^{TT} ({\Bell}_1, {\Bell}_2),
\label{eq:norm}
\end{equation}
with
\begin{equation}
f^{TT} ({\Bell}_1, {\Bell}_2) = [{\Bell} \cdot {\Bell}_1]\tilde{C}^{TT}_{l_1} + [{\Bell} \cdot {\Bell}_2]\tilde{C}^{TT}_{l_2}
\end{equation}
and ${\Bell}={\Bell_1} + {\Bell_2}$.

The mean of each reconstructed convergence stamp is set to zero to remove fluctuations on scales larger than the size of the stamp.  Each reconstructed convergence stamp is then low-pass filtered  by setting modes with $l> 5782$ to zero.  This corresponds to ignoring modes  smaller than the 1.4' beam scale.

The reconstructed lensing convergence stamps from a given ACTPol patch are then stacked  (i.e., averaged).  A `mean field' stamp needs to be subtracted from this stack since the apodization window does not leave the mean of the reconstructed stack identically zero in the absence of any signal \cite{hanson2009,namikawa2013}.  We construct a mean field stamp from the average reconstruction of 15 realizations of random positions in the corresponding ACTPol patch. Each random-position-realization has the same number of stamps as are in the galaxy stack. Thus, by construction, the mean-field-subtracted galaxy stacks show any excess signal above that from random locations.

\begin{figure*}[th!]
\begin{center}
\hspace{-2mm}\includegraphics[width=1.0\columnwidth]{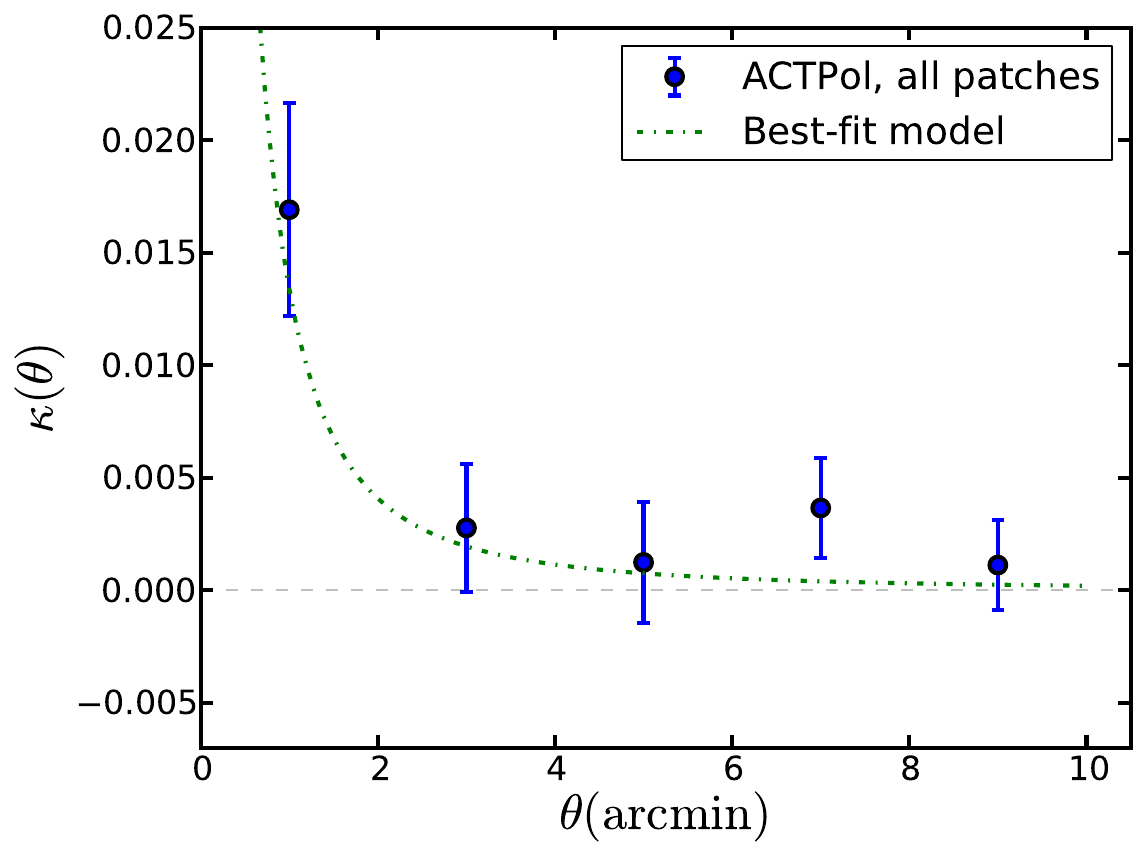}
\hspace{5mm}\includegraphics[width=0.98\columnwidth]{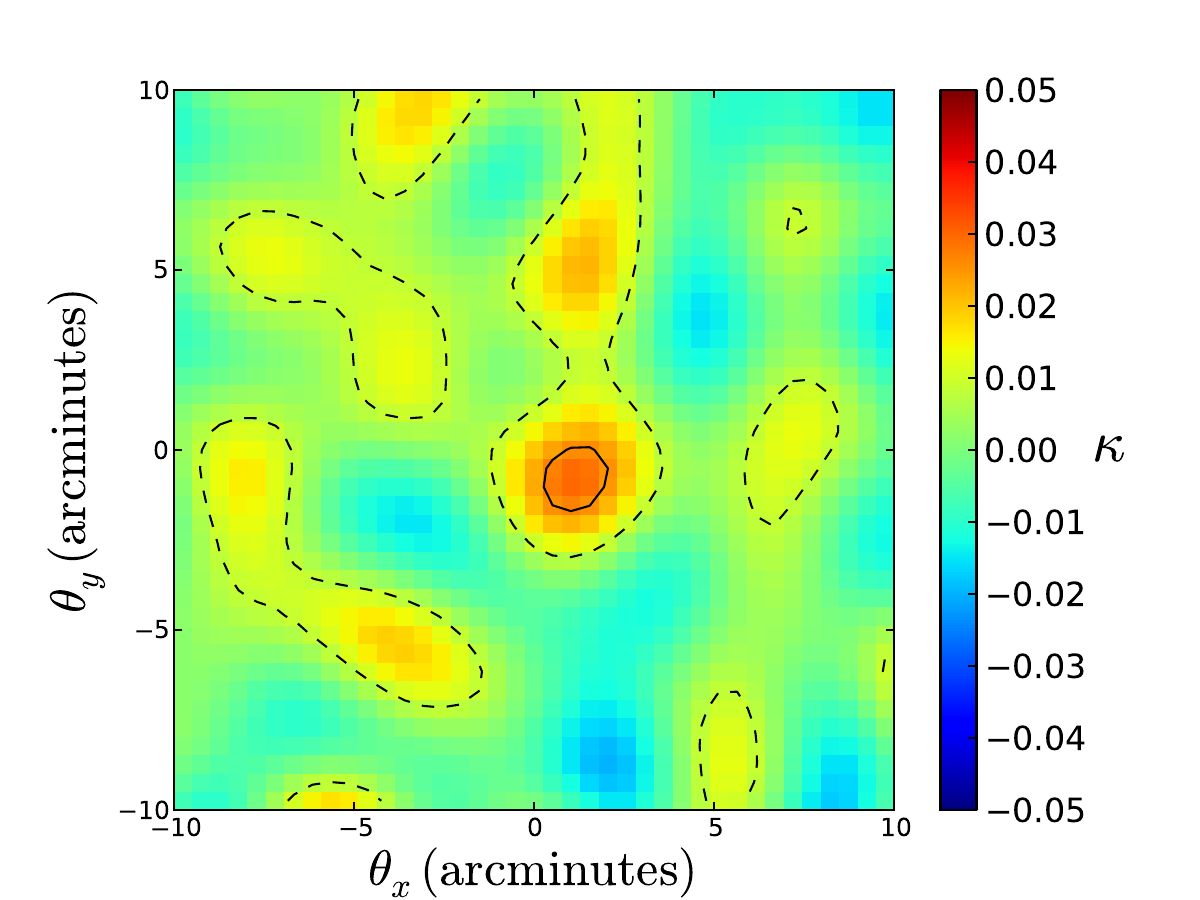}
\caption{{\it Left}: The azimuthally averaged signal from stacked reconstructed convergence stamps centered on CMASS galaxy positions for all three ACTPol deep fields combined.  The green dashed curve shows the best-fit NFW profile.  {\it Right}: The reconstructed convergence stack in the two-dimensional plane, where the horizontal and vertical scales are in arcminutes.  We also show $1\sigma$ (dashed) and $3\sigma$ (solid) contours; the signal is the dark red spot in the middle. The peak is offset by about $1'$ from the center; offsets of $> 1'$ are seen roughly 20\% of the time in simulations of centered input halos given ACTPol noise levels. The detection significance above null is \SigAllNullTen$\sigma$ within 10 arcminutes, and the best-fit curve from \cite{miyatake2013weak} is preferred over null with a significance of \SigAllTheoryTen$\sigma$ within 10 arcminutes. \vspace{-1mm}}
\label{fig:AllPatches}
\end{center}
\end{figure*}

In order to construct the covariance matrix for each patch, we construct 50 independent realizations of simulated ACTPol data for each patch.  These simulations have noise and beam properties matched to the data and include only lensing by large-scale structure.  We repeat the procedure performed on the data on each of the 50 independent simulations. The covariance matrix for each patch is then obtained by calculating the covariance of radial profiles across these 50 mean-field-subtracted, mean stamps. In this way, the covariance matrices capture the correlations between radial bins. This procedure also takes into account any additional covariance coming from overlapping stamps. In addition, it also folds in the uncertainty in the subtracted mean field.\footnote{Note that we use simulations to characterize the covariance matrix since stacking on random positions in the data does not capture the variance due to overlapping stamps and meanfield subtraction. A typical mean-field amplitude is $0.03$, and the uncertainty is $\approx 20\%$ of the errorbars shown in Figure \ref{fig:AllPatches}.}

The pipeline described above is implemented for each ACTPol patch separately as well as for all the patches combined.  The latter is done by stacking the three mean-field-subtracted galaxy stacks for each ACTPol data patch.  The combined-patch covariance matrix is obtained by combining the 50 mean simulated convergence stamps for each patch, and calculating the variance across all 150 mean stamps.  

This pipeline is tested on a suite of simulations where $70' \times 70'$ CMB stamps are lensed with Navarro-Frenk-White (NFW) cluster profiles \cite{navarro1997universal} with varying levels of instrument noise, beam resolution, and pixelization. The pipeline returns unbiased reconstructions (to $\approx 0.1 \sigma$) and S/N estimates in agreement with previous analyses \cite{hu2007cluster}. In particular, the expected detection significance stacking a sample of roughly 12,000 galaxies in lensed CMB stamps with ACTPol beam and noise properties is \PredictedSigAllNullTen$\sigma$.  For this estimate, the masses, concentrations, and redshifts of the lensing galaxies are assumed to be the mean values of the CMASS subsample with optical weak lensing follow up described above \cite{miyatake2013weak}.

\begin{figure}[th!]
\hspace{-2mm}\includegraphics[width=1.0\columnwidth]{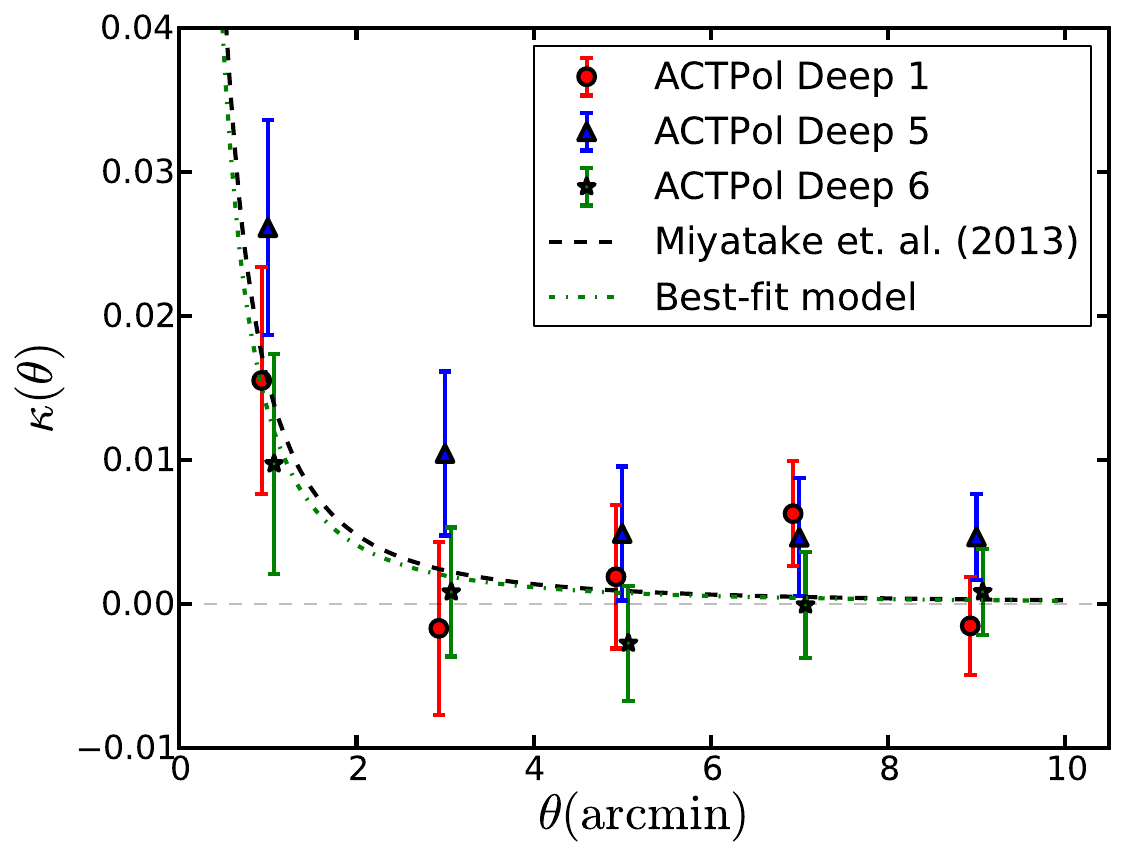}
\caption{Shown are reconstructed convergence profiles centered on CMASS galaxy positions for each ACTPol deep field separately.  The significance with respect to null within 4 arcminutes is \SigDeepOneFour$\sigma$, \SigDeepFiveFour$\sigma$, and \SigDeepSixFour$\sigma$ for ACTPol Deep 1, 5, and 6 respectively.  The green dashed curve is the best-fit NFW profile from all the Deep fields combined, and the black dashed curve is the best-fit NFW profile from a subset of the CMASS galaxies measured via optical weak lensing~\cite{miyatake2013weak}.}
\label{fig:SeparatePatches}
\vspace{-3mm}
\end{figure}

\section{Results}
\label{sec:results}

\begin{figure}[t!]
\begin{center}
\hspace{3mm}
\hspace{-2mm}\includegraphics[width=1.0\columnwidth]{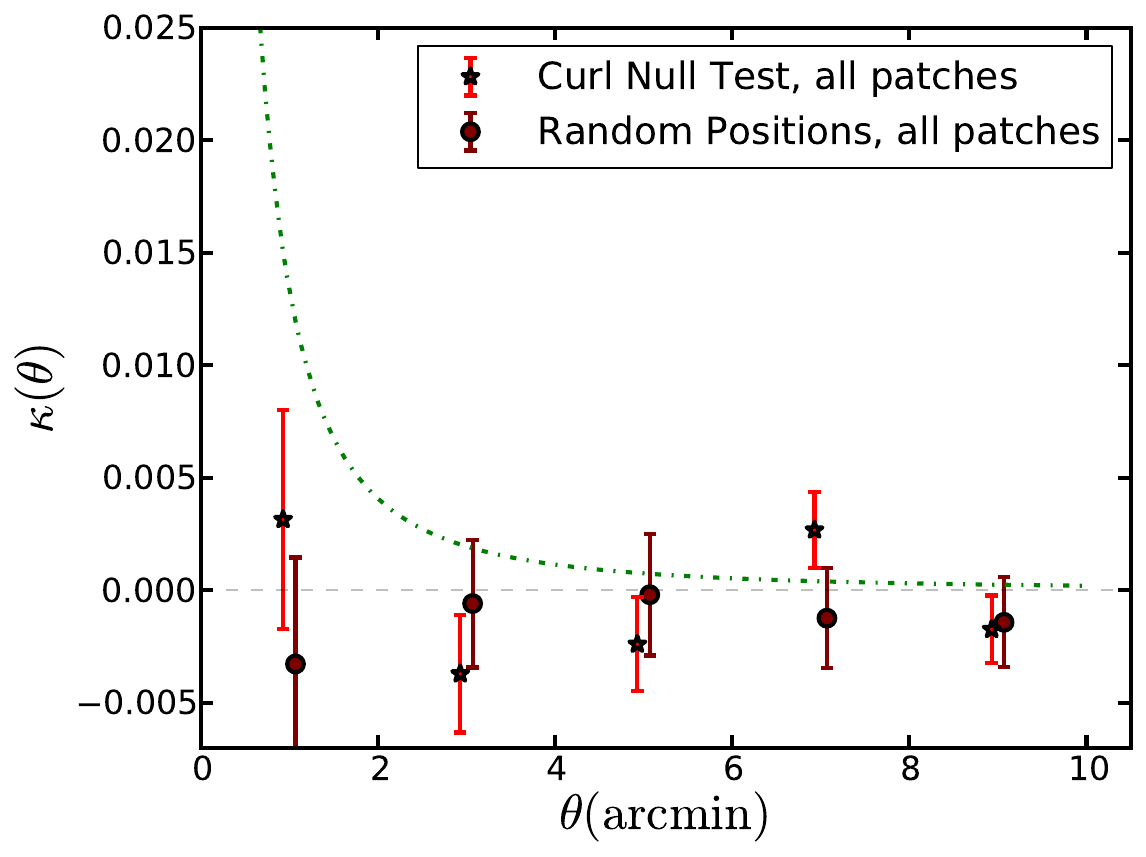}\\
\includegraphics[width=1.0\columnwidth]{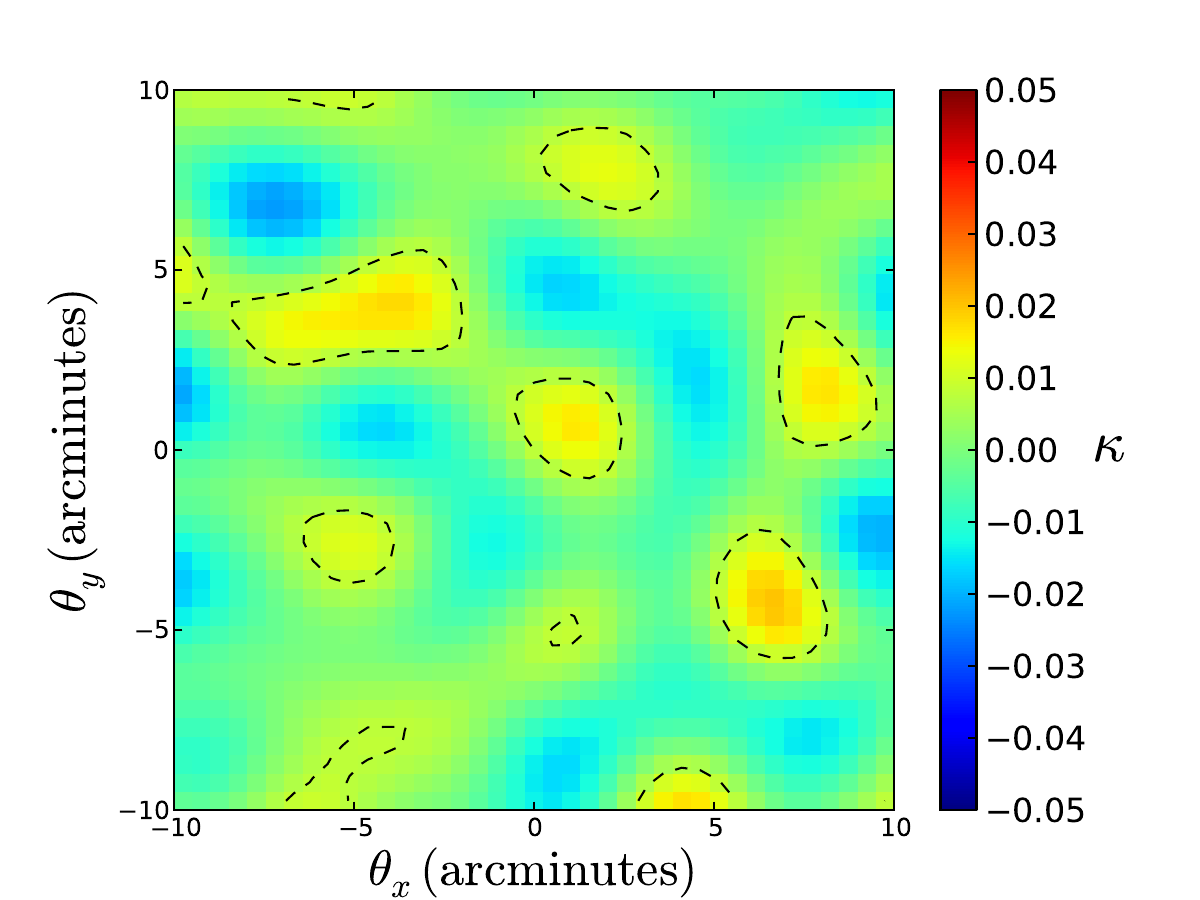}\\
\includegraphics[width=1.0\columnwidth]{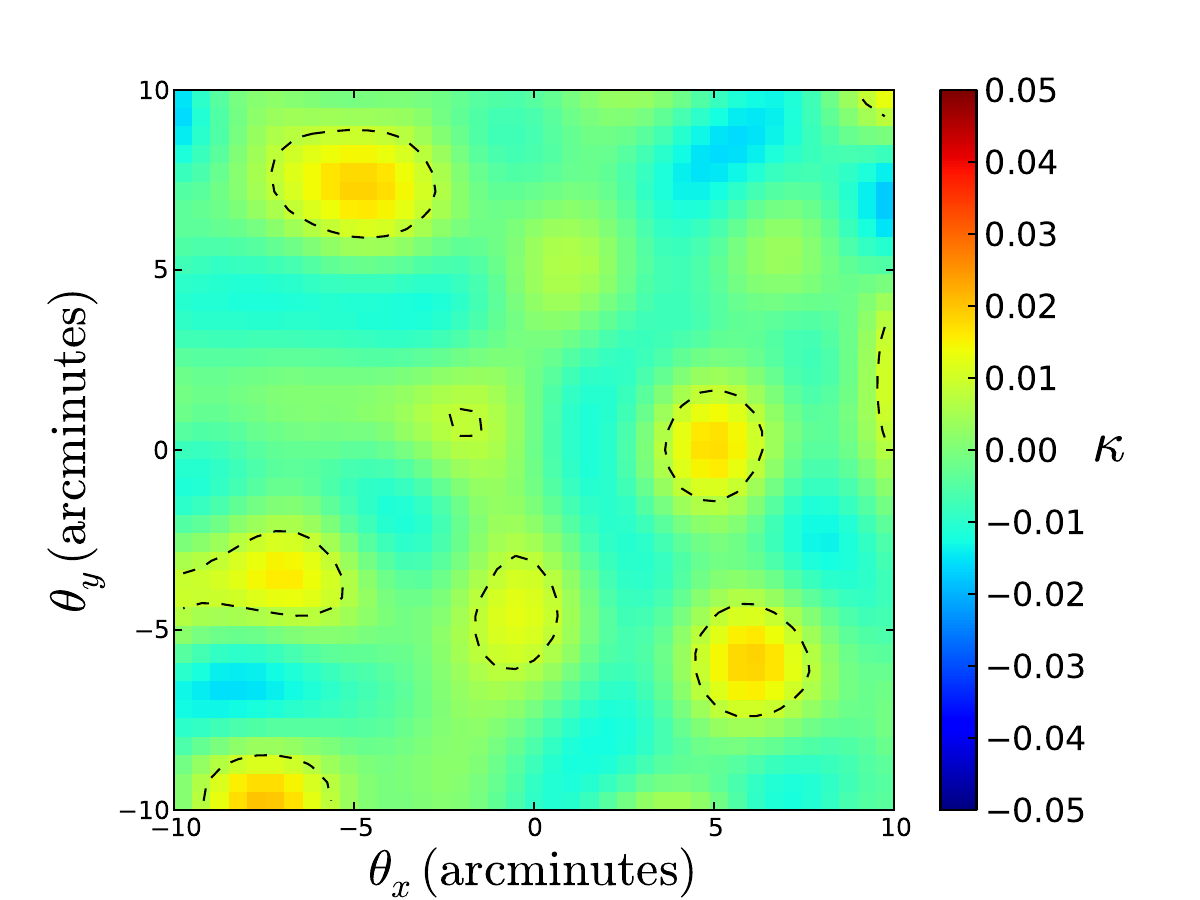}
\caption{{\it Top panel}: Shown are the curl null test performed on the stack of reconstructed convergence stamps centered on CMASS galaxy positions, and a random-position null test where reconstructed convergence stamps are centered on random positions in the data.  {\it Middle and bottom panels}: Shown are the curl and random-position null tests, respectively, in the two-dimensional plane. We also show 1-sigma contours; the lack of a red spot in the middle confirms the null test.}
\label{fig:Null}
\end{center}
 \vspace{-9mm}
\end{figure}

We show the result of the combined-patch stack of reconstructed convergence stamps centered on CMASS galaxies in Figure~\ref{fig:AllPatches}. The left panel shows the measured azimuthally averaged lensing convergence profile, and the right panel shows the reconstructed lensing stack in the two-dimensional plane.  We note that the signal peak in the two-dimensional plot is offset by about $1'$. This is also seen in simulations of centered input halos given ACTPol noise levels, where offsets of $> 1'$ are seen roughly 20\% of the time.  We also note that this offset is well within the virial radius of CMASS halos. The profile has been binned, with inverse-variance weighting, in annuli that are four-pixels (2 arcminutes) wide so that correlations between neighboring bins in general do not exceed 50\%. The exceptions are that for the stacks on galaxy positions, the 3rd and 4th bins are correlated by 65\% and the 4th and 5th bins are correlated by 70\%. This is due to overlapping stamps, as the galaxy locations are more correlated than random positions.  

The significance of this detection above the null hypothesis, including measured points within 10 arcminutes of the profile center, is \SigAllNullTen$\sigma$.  This is calculated using the combined-patch covariance matrix, $\bf C$, where

\begin{equation}
\Big(\frac{S}{N}\Big)^2 = \chi^2_{\rm null}=\sum_{\theta_1, \theta_2 \leq 10'} \kappa(\theta_1) {\bf {C}}^{-1} \kappa(\theta_2).
\end{equation}
Restricting this to 4 arcminutes from the profile center, where most of the S/N is from, gives a detection significance above null of \SigAllNullFour$\sigma$.

We fit the data points within 10 arcminutes from the center with an NFW profile, which is the projected and redshift-averaged mass density as in, e.g., \cite{Bartlemann96}. We vary the mass and concentration and obtain a best-fit profile with a mass of \bestfitmass~and a concentration of \bestfitconcentration. This result is obtained by imposing a prior on the c-M relation from~\cite{Maccio2007} assuming Gaussian errors on the normalization of this relation of 20\% as found in~\cite{miyatake2013weak}.  We note that the best-fit mass and mass error are unchanged with and without the prior; however, since there is significant degeneracy in the concentration, given our noise levels, the prior influences the best-fit $c_{200\bar{\rho}_0}$ and corresponding error.  This best-fit curve gives a reduced chi-square of \bestfitgoodness~for $\nu=3$ degrees of freedom, and is consistent with the best-fit curve from \cite{miyatake2013weak}.   The data also favors the best-fit curve from \cite{miyatake2013weak} over the null line ($\kappa=0$) at a significance of  \SigAllTheoryTen$\sigma$ within 10 arcminutes, where we calculate this significance using $\sqrt{\chi^2_{{\rm null}} - \chi^2_{{\rm best-fit}} }$.  Restricting to within 4 arcminutes, the model is favored over null with a significance of \SigAllTheoryFour$\sigma$.

The profile of the reconstructed lensing stack for each ACTPol patch is shown in Figure~\ref{fig:SeparatePatches}. An excess above null is seen in all three patches with a significance of \SigDeepOneFour$\sigma$, \SigDeepFiveFour$\sigma$, and \SigDeepSixFour$\sigma$ within 4 arcminutes for D1, D5, and D6 respectively.  
The black-dashed curve in Figure~\ref{fig:SeparatePatches} is an NFW profile with the best-fit mass and concentration found from optical weak lensing of a subset of the CMASS galaxy sample \cite{miyatake2013weak}.  This best-fit mass and concentration for the subset is $M_{200\bar{\rho}_0} = 2.3 \times 10^{13}~h^{-1} M_{\odot}$ and $c_{200\bar{\rho}_0} = 5.0$, where the concentration is from the best-fit concentration-mass relation found in~\cite{miyatake2013weak}, calculated at the mean redshift of the subset ($z=0.55$).\footnote{In \cite{miyatake2013weak}, a best-fit of $c_{200\bar{\rho}_0} = 5.0$ is found for CMASS galaxies when their model allows for off-centering of CMASS galaxies in dark matter halos.  Without this degree of freedom, a best-fit of $c_{200\bar{\rho}_0} = 3.2$ is found.}

\section{Systematic Checks}
\label{sec:checks}

Two different null tests are performed to verify the robustness of the signal. The first is to stack on random positions in the data.  As mentioned above, all of the stacked images have a subtracted mean field stamp that is determined from averaging 15 realizations of randomly selected stamps from the data.  Therefore, by construction the measured signal is the excess above that from random locations. However, we show a single random-position realization which contains the same number of stamps as are in the galaxy stack.  We subtract the mean field stamp from this single realization and plot the resulting profile in the top panel of Figure \ref{fig:Null} (brown circles).  The data points are consistent with the null hypothesis with a probability-to-exceed (PTE) of 0.92.

The second null test is a curl test where we repeat the analysis of stacking reconstructions centered on CMASS galaxies and subtract a mean field stamp as before.  However, this time the divergence in Eq \ref{eq:main} is replaced with a curl, and the first instance of the dot product ${\Bell} \cdot {\Bell}_1$ in Eq \ref{eq:norm} (not in $f^{TT}$) is replaced with a cross product \cite{cooray2005,sherwin2012,vanengelen/etal/2012}, where both the curl and cross product are projected perpendicular to the image plane. The reconstruction is then expected to contain only noise since lensing is not expected to generate a curl signal in temperature maps. The curl reconstruction data points scatter about zero, with a PTE of 0.08, as shown in Figure \ref{fig:Null} (red stars).

As can be seen in Figure \ref{fig:SeparatePatches}, the mean signal is highest in D5. A histogram analysis of the stamps in both D5 and in the quadrant of D5 with the highest mean signal shows no apparent outliers. We note that excluding this quadrant from our analysis still results in a S/N $>3\sigma$ within 10 arcminutes.

We also consider several possible contaminants that could bias a detection of CMB halo lensing. Ionized gas in clusters hosting the stacked galaxies could produce a decrement in the CMB temperature at 146 GHz due to the thermal Sunyaev-Zeldovich (tSZ) effect \cite{SZ1970,SZ1972}. In order to determine the effect of such a contaminant on the lensing reconstruction, we added a Gaussian decrement with a peak value of $-35 \mu K$ and 1$\sigma$ width of 1 arcminute\footnote{The virial radius of a $10^{13}M_\odot$ halo at $z=0.6$ is roughly $1.5'$.} to CMB temperature maps lensed by NFW profiles as discussed above. We adopted this as a conservative level of tSZ for CMASS halos (see for example \cite{HandSZ}).  This contamination resulted in the reconstruction being biased low by about $0.3\sigma$ within 3 arcminutes at ACTPol noise levels, with negligible bias beyond 3 arcminutes. An identical check was performed for $35 \mu K$ increments (corresponding to point source emission) with a similar suppression of the signal.  In addition, no appreciable tSZ decrement or point source increment is found when stacking the stamps taken directly from CMB temperature maps and centered on the CMASS galaxies, after these stamps have been filtered to isolate modes between $1000 < l < 8000$. These checks indicate that the detected positive signals in Figures \ref{fig:AllPatches} and \ref{fig:SeparatePatches} do not arise from tSZ or point source emission. The kinetic SZ effect due to the bulk motion of the cluster will produce a similar symmetric increment or decrement. Furthermore, asymmetric contaminants, like those due to the kinetic SZ effect from internal gas motions, do not coherently align with the CMB gradient and only add noise by construction of the estimator. 

The stacked lensing convergence measured in Figures \ref{fig:AllPatches} and \ref{fig:SeparatePatches} could also have contributions that are not due to CMB lensing by the halo that each galaxy resides in (the 1-halo term), but instead are due to correlated halos in the vicinity of the galaxies (the 2-halo term, \cite{MaFry2Halo,Seljak2Halo}). Since most of our detected signal is within a 2 arcminute region, where the 1-halo term dominates over the 2-halo term (see for example Figure 7 in \cite{miyatake2013weak}), one would not expect the 2-halo term to contribute significantly to the detection significance in this work.

\section{Discussion}
\label{sec:discussion}

We have presented the stacked reconstructed lensing convergence of CMASS galaxies within the first season ACTPol deep fields and shown evidence of CMB lensing from these halos at a significance of \SigAllNullTen$\sigma$ above null. The lensing convergence is directly related to the projected density profile of these halos and hence our results demonstrate that it is possible to constrain the mass profile of massive objects using CMB lensing alone.

We find a best-fit mass and concentration from the stacked convergence stamps of \bestfitmass~and \bestfitconcentration~fitting to an NFW profile.  These mass and concentration values are in broad agreement with the optical weak lensing estimates in \cite{miyatake2013weak} based on a subset of the CMASS galaxy sample.   Our data also favors the best-fit profile from \cite{miyatake2013weak} over a null line at a significance of \SigAllTheoryTen$\sigma$ within 10 arcminutes. 

With this work we demonstrate that CMB observations are now achieving the sensitivity and resolution to provide mass estimates of dark matter halos belonging to galaxy groups and clusters.  With the advent of next-generation CMB surveys, we expect this technique to be further exploited, thus opening a new window on the dark Universe.\\ \\

\begin{acknowledgments}
The authors would like to thank Hironao Miyatake, Surhud More, and Anze Slosar for useful discussions regarding CMASS 
and BOSS galaxies.  MM acknowledges support from an SBU-BNL Research Initiatives Seed Grant: Award Number 37298, Project Number 1111593.
This work was supported by the U.S. National Science Foundation through awards
AST-0408698 and AST-0965625 for the ACT project, as well as awards PHY-0855887
and PHY-1214379. Funding was also provided by Princeton University, the
University of Pennsylvania, Cornell University, and a Canada Foundation for Innovation (CFI) award
to UBC. ACT operates in the Parque Astron\'omico Atacama in northern Chile
under the auspices of the Comisi\'on Nacional de Investigaci\'on Cient\'ifica y
Tecnol\'ogica de Chile (CONICYT). Computations were performed on the GPC
supercomputer at the SciNet HPC Consortium. SciNet is funded by the CFI under
the auspices of Compute Canada, the Government of Ontario, the Ontario Research
Fund -- Research Excellence; and the University of Toronto. The development of
multichroic detectors and lenses was supported by NASA grants NNX13AE56G and
NNX14AB58G. Funding from ERC grant 259505 supports SN, JD, and TL. RD was supported by CONICYT grants QUIMAL-120001 and FONDECYT-1141113. We gratefully acknowledge support from the Misrahi and Wilkinson
research funds. 

\end{acknowledgments}

\bibliography{refs.bib,lens_refs,apj-jour}

\end{document}

%% file: authors.tex
\author{Mathew~Madhavacheril}
\affiliation{Physics and Astronomy Department, Stony Brook University, Stony Brook, NY USA 11794}
\author{Neelima~Sehgal}
\affiliation{Physics and Astronomy Department, Stony Brook University, Stony Brook, NY USA 11794}
\author{Rupert~Allison}
\affiliation{Sub-Department of Astrophysics, University of Oxford, Keble Road, Oxford, UK OX1 3RH}
\author{Nick~Battaglia}
\affiliation{McWilliams Center for Cosmology, Carnegie Mellon University, Department of Physics, 5000 Forbes Ave., Pittsburgh PA, USA, 15213}
\author{J~Richard~Bond}
\affiliation{Canadian Institute for Theoretical Astrophysics, University of
Toronto, Toronto, ON, Canada M5S 3H8}
\author{Erminia~Calabrese}
\affiliation{Sub-Department of Astrophysics, University of Oxford, Keble Road, Oxford, UK OX1 3RH}
\author{Jerod~Caligiuri}
\affiliation{Department of Physics and Astronomy, University of Pittsburgh, 
Pittsburgh, PA, USA 15260}
\author{Kevin~Coughlin}
\affiliation{Department of Physics, University of Michigan, Ann Arbor, USA 48103}
\author{Devin~Crichton}
\affiliation{Dept. of Physics and Astronomy, The Johns Hopkins University, 3400 N. Charles St., Baltimore, MD, USA 21218-2686}
\author{Rahul~Datta}
\affiliation{Department of Physics, University of Michigan, Ann Arbor, USA 48103}
\author{Mark~J.~Devlin}
\affiliation{Department of Physics and Astronomy, University of
Pennsylvania, 209 South 33rd Street, Philadelphia, PA, USA 19104}
\author{Joanna~Dunkley}
\affiliation{Sub-Department of Astrophysics, University of Oxford, Keble Road, Oxford, UK OX1 3RH}
\author{Rolando~D\"{u}nner}
\affiliation{Departamento de Astronom{\'{i}}a y Astrof{\'{i}}sica, Pontific\'{i}a Universidad Cat\'{o}lica,
Casilla 306, Santiago 22, Chile}
\author{Kevin~Fogarty}
\affiliation{Dept. of Physics and Astronomy, The Johns Hopkins University, 3400 N. Charles St., Baltimore, MD, USA 21218-2686}
\author{Emily~Grace}
\affiliation{Joseph Henry Laboratories of Physics, Jadwin Hall,
Princeton University, Princeton, NJ, USA 08544}
\author{Amir~Hajian}
\affiliation{Canadian Institute for Theoretical Astrophysics, University of
Toronto, Toronto, ON, Canada M5S 3H8}
\author{Matthew~Hasselfield}
\affiliation{Department of Astrophysical Sciences, Peyton Hall, 
Princeton University, Princeton, NJ USA 08544}
\author{J.~Colin~Hill}
\affiliation{Dept. of Astronomy, Pupin Hall, Columbia University, New
York, NY USA 10027}
\author{Matt~Hilton}
\affiliation{Astrophysics and Cosmology Research Unit, School of Mathematics, Statistics and Computer Science, University of KwaZulu-Natal, Durban 4041, South Africa}

\author{Adam~D.~Hincks}
\affiliation{Department of Physics and Astronomy, University of
British Columbia, Vancouver, BC, Canada V6T 1Z4}
\author{Ren\'ee~Hlozek}
\affiliation{Department of Astrophysical Sciences, Peyton Hall, 
Princeton University, Princeton, NJ USA 08544}
\author{John~P.~Hughes}
\affiliation{Department of Physics and Astronomy, Rutgers, 
The State University of New Jersey, Piscataway, NJ USA 08854-8019}
\author{Arthur~Kosowsky}
\affiliation{Department of Physics and Astronomy, University of Pittsburgh, 
Pittsburgh, PA, USA 15260}
\author{Thibaut~Louis}
\affiliation{Sub-Department of Astrophysics, University of Oxford, Keble Road, Oxford, UK OX1 3RH}
\author{Marius~Lungu}
\affiliation{Department of Physics and Astronomy, University of
Pennsylvania, 209 South 33rd Street, Philadelphia, PA, USA 19104}
\author{Jeff~McMahon}
\affiliation{Department of Physics, University of Michigan, Ann Arbor, USA 48103}
\author{Kavilan~Moodley}
\affiliation{Astrophysics and Cosmology Research Unit, School of Mathematics, Statistics and Computer Science, University of KwaZulu-Natal, Durban 4041, South Africa}
\author{Charles~Munson}
\affiliation{Department of Physics, University of Michigan, Ann Arbor, USA 48103}
\author{Sigurd~Naess}
\affiliation{Sub-Department of Astrophysics, University of Oxford, Keble Road, Oxford, UK OX1 3RH}

\author{Federico~Nati}
\affiliation{Dipartimento di Fisica, Universit\`{a} La Sapienza, P. le A. Moro 2, 00185 Roma, Italy}
\author{Laura~Newburgh}
\affiliation{Dunlap Institute, University of Toronto, 50 St. George St. Toronto ON M5S 3H4}
\author{Michael~D.~Niemack}
\affiliation{Department of Physics, Cornell University, Ithaca, NY, USA 14853}
\author{Lyman~A.~Page}
\affiliation{Joseph Henry Laboratories of Physics, Jadwin Hall,
Princeton University, Princeton, NJ, USA 08544}
\author{Bruce~Partridge}
\affiliation{Department of Physics and Astronomy, Haverford College,
Haverford, PA, USA 19041}
\author{Benjamin~Schmitt}
\affiliation{Department of Physics and Astronomy, University of
Pennsylvania, 209 South 33rd Street, Philadelphia, PA, USA 19104}
\author{Blake~D.~Sherwin}
\affiliation{Berkeley Center for Cosmological Physics, LBL and
Department of Physics, University of California, Berkeley, CA, USA 94720}
\author{Jon~Sievers}
\affiliation{Astrophysics and Cosmology Research Unit, School of Chemistry and Physics, University of KwaZulu-Natal, Durban 4041, South Africa}
\affiliation{National Institute for Theoretical Physics (NITheP), University of KwaZulu-Natal, Private Bag X54001, Durban 4000, South Africa}
\author{David~N.~Spergel}
\affiliation{Department of Astrophysical Sciences, Peyton Hall, 
Princeton University, Princeton, NJ USA 08544}
\author{Suzanne~T.~Staggs}
\affiliation{Joseph Henry Laboratories of Physics, Jadwin Hall,
Princeton University, Princeton, NJ, USA 08544}
\author{Robert~Thornton}
\affiliation{Department of Physics , West Chester University 
of Pennsylvania, West Chester, PA, USA 19383}
\affiliation{Department of Physics and Astronomy, University of
Pennsylvania, 209 South 33rd Street, Philadelphia, PA, USA 19104}
\author{Alexander~Van~Engelen}
\affiliation{Canadian Institute for Theoretical Astrophysics, University of
Toronto, Toronto, ON, Canada M5S 3H8}
\author{Jonathan~T.~Ward}
\affiliation{Department of Physics and Astronomy, University of
Pennsylvania, 209 South 33rd Street, Philadelphia, PA, USA 19104}
\author{Edward~J.~Wollack}
\affiliation{NASA/Goddard Space Flight Center, Greenbelt, MD, USA 20771}

%% file: ClusterLens.bbl
\begin{thebibliography}{62}
\expandafter\ifx\csname natexlab\endcsname\relax\def\natexlab#1{#1}\fi
\expandafter\ifx\csname bibnamefont\endcsname\relax
  \def\bibnamefont#1{#1}\fi
\expandafter\ifx\csname bibfnamefont\endcsname\relax
  \def\bibfnamefont#1{#1}\fi
\expandafter\ifx\csname citenamefont\endcsname\relax
  \def\citenamefont#1{#1}\fi
\expandafter\ifx\csname url\endcsname\relax
  \def\url#1{\texttt{#1}}\fi
\expandafter\ifx\csname urlprefix\endcsname\relax\def\urlprefix{URL }\fi
\providecommand{\bibinfo}[2]{#2}
\providecommand{\eprint}[2][]{\url{#2}}

\bibitem[{\citenamefont{{Smith} et~al.}(2007)\citenamefont{{Smith}, {Zahn}, and
  {Dor{\'e}}}}]{smith/etal/2007}
\bibinfo{author}{\bibfnamefont{K.~M.} \bibnamefont{{Smith}}},
  \bibinfo{author}{\bibfnamefont{O.}~\bibnamefont{{Zahn}}}, \bibnamefont{and}
  \bibinfo{author}{\bibfnamefont{O.}~\bibnamefont{{Dor{\'e}}}},
  \bibinfo{journal}{\prd} \textbf{\bibinfo{volume}{76}},
  \bibinfo{pages}{043510} (\bibinfo{year}{2007}), \eprint{0705.3980}.

\bibitem[{\citenamefont{{Hirata} et~al.}(2008)\citenamefont{{Hirata}, {Ho},
  {Padmanabhan}, {Seljak}, and {Bahcall}}}]{hirata/etal/2008}
\bibinfo{author}{\bibfnamefont{C.~M.} \bibnamefont{{Hirata}}},
  \bibinfo{author}{\bibfnamefont{S.}~\bibnamefont{{Ho}}},
  \bibinfo{author}{\bibfnamefont{N.}~\bibnamefont{{Padmanabhan}}},
  \bibinfo{author}{\bibfnamefont{U.}~\bibnamefont{{Seljak}}}, \bibnamefont{and}
  \bibinfo{author}{\bibfnamefont{N.~A.} \bibnamefont{{Bahcall}}},
  \bibinfo{journal}{\prd} \textbf{\bibinfo{volume}{78}}, \bibinfo{eid}{043520}
  (\bibinfo{year}{2008}), \eprint{0801.0644}.

\bibitem[{\citenamefont{{Das} et~al.}(2011)\citenamefont{{Das}, {Marriage},
  {Ade}, {Aguirre}, {Amiri}, {Appel}, {Barrientos}, {Battistelli}, {Bond},
  {Brown} et~al.}}]{das/etal/2011}
\bibinfo{author}{\bibfnamefont{S.}~\bibnamefont{{Das}}},
  \bibinfo{author}{\bibfnamefont{T.~A.} \bibnamefont{{Marriage}}},
  \bibinfo{author}{\bibfnamefont{P.~A.~R.} \bibnamefont{{Ade}}},
  \bibinfo{author}{\bibfnamefont{P.}~\bibnamefont{{Aguirre}}},
  \bibinfo{author}{\bibfnamefont{M.}~\bibnamefont{{Amiri}}},
  \bibinfo{author}{\bibfnamefont{J.~W.} \bibnamefont{{Appel}}},
  \bibinfo{author}{\bibfnamefont{L.~F.} \bibnamefont{{Barrientos}}},
  \bibinfo{author}{\bibfnamefont{E.~S.} \bibnamefont{{Battistelli}}},
  \bibinfo{author}{\bibfnamefont{J.~R.} \bibnamefont{{Bond}}},
  \bibinfo{author}{\bibfnamefont{B.}~\bibnamefont{{Brown}}},
  \bibnamefont{et~al.}, \bibinfo{journal}{\apj} \textbf{\bibinfo{volume}{729}},
  \bibinfo{eid}{62} (\bibinfo{year}{2011}), \eprint{1009.0847}.

\bibitem[{\citenamefont{{van Engelen} et~al.}(2012)\citenamefont{{van Engelen},
  {Keisler}, {Zahn}, {Aird}, {Benson}, {Bleem}, {Carlstrom}, {Chang}, {Cho},
  {Crawford} et~al.}}]{vanengelen/etal/2012}
\bibinfo{author}{\bibfnamefont{A.}~\bibnamefont{{van Engelen}}},
  \bibinfo{author}{\bibfnamefont{R.}~\bibnamefont{{Keisler}}},
  \bibinfo{author}{\bibfnamefont{O.}~\bibnamefont{{Zahn}}},
  \bibinfo{author}{\bibfnamefont{K.~A.} \bibnamefont{{Aird}}},
  \bibinfo{author}{\bibfnamefont{B.~A.} \bibnamefont{{Benson}}},
  \bibinfo{author}{\bibfnamefont{L.~E.} \bibnamefont{{Bleem}}},
  \bibinfo{author}{\bibfnamefont{J.~E.} \bibnamefont{{Carlstrom}}},
  \bibinfo{author}{\bibfnamefont{C.~L.} \bibnamefont{{Chang}}},
  \bibinfo{author}{\bibfnamefont{H.~M.} \bibnamefont{{Cho}}},
  \bibinfo{author}{\bibfnamefont{T.~M.} \bibnamefont{{Crawford}}},
  \bibnamefont{et~al.}, \bibinfo{journal}{\apj} \textbf{\bibinfo{volume}{756}},
  \bibinfo{eid}{142} (\bibinfo{year}{2012}), \eprint{1202.0546}.

\bibitem[{\citenamefont{{Planck
  Collaboration}}(2014{\natexlab{a}})}]{planck_lens/2013}
\bibinfo{author}{\bibnamefont{{Planck Collaboration}}}, \bibinfo{journal}{\aap}
  \textbf{\bibinfo{volume}{571}}, \bibinfo{eid}{A17}
  (\bibinfo{year}{2014}{\natexlab{a}}), \eprint{1303.5077}.

\bibitem[{\citenamefont{{Sherwin} et~al.}(2012)\citenamefont{{Sherwin}, {Das},
  {Hajian}, {Addison}, {Bond}, {Crichton}, {Devlin}, {Dunkley}, {Gralla},
  {Halpern} et~al.}}]{sherwin/etal/2012}
\bibinfo{author}{\bibfnamefont{B.~D.} \bibnamefont{{Sherwin}}},
  \bibinfo{author}{\bibfnamefont{S.}~\bibnamefont{{Das}}},
  \bibinfo{author}{\bibfnamefont{A.}~\bibnamefont{{Hajian}}},
  \bibinfo{author}{\bibfnamefont{G.}~\bibnamefont{{Addison}}},
  \bibinfo{author}{\bibfnamefont{J.~R.} \bibnamefont{{Bond}}},
  \bibinfo{author}{\bibfnamefont{D.}~\bibnamefont{{Crichton}}},
  \bibinfo{author}{\bibfnamefont{M.~J.} \bibnamefont{{Devlin}}},
  \bibinfo{author}{\bibfnamefont{J.}~\bibnamefont{{Dunkley}}},
  \bibinfo{author}{\bibfnamefont{M.~B.} \bibnamefont{{Gralla}}},
  \bibinfo{author}{\bibfnamefont{M.}~\bibnamefont{{Halpern}}},
  \bibnamefont{et~al.}, \bibinfo{journal}{\prd} \textbf{\bibinfo{volume}{86}},
  \bibinfo{eid}{083006} (\bibinfo{year}{2012}), \eprint{1207.4543}.

\bibitem[{\citenamefont{{Bleem} et~al.}(2012)\citenamefont{{Bleem}, {van
  Engelen}, {Holder}, {Aird}, {Armstrong}, {Ashby}, {Becker}, {Benson},
  {Biesiadzinski}, {Brodwin} et~al.}}]{bleem/etal/2012}
\bibinfo{author}{\bibfnamefont{L.~E.} \bibnamefont{{Bleem}}},
  \bibinfo{author}{\bibfnamefont{A.}~\bibnamefont{{van Engelen}}},
  \bibinfo{author}{\bibfnamefont{G.~P.} \bibnamefont{{Holder}}},
  \bibinfo{author}{\bibfnamefont{K.~A.} \bibnamefont{{Aird}}},
  \bibinfo{author}{\bibfnamefont{R.}~\bibnamefont{{Armstrong}}},
  \bibinfo{author}{\bibfnamefont{M.~L.~N.} \bibnamefont{{Ashby}}},
  \bibinfo{author}{\bibfnamefont{M.~R.} \bibnamefont{{Becker}}},
  \bibinfo{author}{\bibfnamefont{B.~A.} \bibnamefont{{Benson}}},
  \bibinfo{author}{\bibfnamefont{T.}~\bibnamefont{{Biesiadzinski}}},
  \bibinfo{author}{\bibfnamefont{M.}~\bibnamefont{{Brodwin}}},
  \bibnamefont{et~al.}, \bibinfo{journal}{\apj} \textbf{\bibinfo{volume}{753}},
  \bibinfo{pages}{L9} (\bibinfo{year}{2012}), \eprint{1203.4808}.

\bibitem[{\citenamefont{{Geach} et~al.}(2013)\citenamefont{{Geach}, {Hickox},
  {Bleem}, {Brodwin}, {Holder}, {Aird}, {Benson}, {Bhattacharya}, {Carlstrom},
  {Chang} et~al.}}]{geach/etal/2013}
\bibinfo{author}{\bibfnamefont{J.~E.} \bibnamefont{{Geach}}},
  \bibinfo{author}{\bibfnamefont{R.~C.} \bibnamefont{{Hickox}}},
  \bibinfo{author}{\bibfnamefont{L.~E.} \bibnamefont{{Bleem}}},
  \bibinfo{author}{\bibfnamefont{M.}~\bibnamefont{{Brodwin}}},
  \bibinfo{author}{\bibfnamefont{G.~P.} \bibnamefont{{Holder}}},
  \bibinfo{author}{\bibfnamefont{K.~A.} \bibnamefont{{Aird}}},
  \bibinfo{author}{\bibfnamefont{B.~A.} \bibnamefont{{Benson}}},
  \bibinfo{author}{\bibfnamefont{S.}~\bibnamefont{{Bhattacharya}}},
  \bibinfo{author}{\bibfnamefont{J.~E.} \bibnamefont{{Carlstrom}}},
  \bibinfo{author}{\bibfnamefont{C.~L.} \bibnamefont{{Chang}}},
  \bibnamefont{et~al.}, \bibinfo{journal}{\apjl}
  \textbf{\bibinfo{volume}{776}}, \bibinfo{eid}{L41} (\bibinfo{year}{2013}),
  \eprint{1307.1706}.

\bibitem[{\citenamefont{{Holder} et~al.}(2013)\citenamefont{{Holder}, {Viero},
  {Zahn}, {Aird}, {Benson}, {Bhattacharya}, {Bleem}, {Bock}, {Brodwin},
  {Carlstrom} et~al.}}]{holder/etal/2013}
\bibinfo{author}{\bibfnamefont{G.~P.} \bibnamefont{{Holder}}},
  \bibinfo{author}{\bibfnamefont{M.~P.} \bibnamefont{{Viero}}},
  \bibinfo{author}{\bibfnamefont{O.}~\bibnamefont{{Zahn}}},
  \bibinfo{author}{\bibfnamefont{K.~A.} \bibnamefont{{Aird}}},
  \bibinfo{author}{\bibfnamefont{B.~A.} \bibnamefont{{Benson}}},
  \bibinfo{author}{\bibfnamefont{S.}~\bibnamefont{{Bhattacharya}}},
  \bibinfo{author}{\bibfnamefont{L.~E.} \bibnamefont{{Bleem}}},
  \bibinfo{author}{\bibfnamefont{J.}~\bibnamefont{{Bock}}},
  \bibinfo{author}{\bibfnamefont{M.}~\bibnamefont{{Brodwin}}},
  \bibinfo{author}{\bibfnamefont{J.~E.} \bibnamefont{{Carlstrom}}},
  \bibnamefont{et~al.}, \bibinfo{journal}{\apjl}
  \textbf{\bibinfo{volume}{771}}, \bibinfo{eid}{L16} (\bibinfo{year}{2013}),
  \eprint{1303.5048}.

\bibitem[{\citenamefont{{Planck
  Collaboration}}(2014{\natexlab{b}})}]{planck_ciblensing/2013}
\bibinfo{author}{\bibnamefont{{Planck Collaboration}}}, \bibinfo{journal}{\aap}
  \textbf{\bibinfo{volume}{571}}, \bibinfo{eid}{A18}
  (\bibinfo{year}{2014}{\natexlab{b}}), \eprint{1303.5078}.

\bibitem[{\citenamefont{{Hanson} et~al.}(2013)\citenamefont{{Hanson}, {Hoover},
  {Crites}, {Ade}, {Aird}, {Austermann}, {Beall}, {Bender}, {Benson}, {Bleem}
  et~al.}}]{hanson/etal/2013}
\bibinfo{author}{\bibfnamefont{D.}~\bibnamefont{{Hanson}}},
  \bibinfo{author}{\bibfnamefont{S.}~\bibnamefont{{Hoover}}},
  \bibinfo{author}{\bibfnamefont{A.}~\bibnamefont{{Crites}}},
  \bibinfo{author}{\bibfnamefont{P.~A.~R.} \bibnamefont{{Ade}}},
  \bibinfo{author}{\bibfnamefont{K.~A.} \bibnamefont{{Aird}}},
  \bibinfo{author}{\bibfnamefont{J.~E.} \bibnamefont{{Austermann}}},
  \bibinfo{author}{\bibfnamefont{J.~A.} \bibnamefont{{Beall}}},
  \bibinfo{author}{\bibfnamefont{A.~N.} \bibnamefont{{Bender}}},
  \bibinfo{author}{\bibfnamefont{B.~A.} \bibnamefont{{Benson}}},
  \bibinfo{author}{\bibfnamefont{L.~E.} \bibnamefont{{Bleem}}},
  \bibnamefont{et~al.}, \bibinfo{journal}{Physical Review Letters}
  \textbf{\bibinfo{volume}{111}}, \bibinfo{eid}{141301} (\bibinfo{year}{2013}),
  \eprint{1307.5830}.

\bibitem[{\citenamefont{{Ade} et~al.}(2014)\citenamefont{{Ade}, {Akiba},
  {Anthony}, {Arnold}, {Atlas}, {Barron}, {Boettger}, {Borrill}, {Borys},
  {Chapman} et~al.}}]{pbear-herschel/2013}
\bibinfo{author}{\bibfnamefont{P.~A.~R.} \bibnamefont{{Ade}}},
  \bibinfo{author}{\bibfnamefont{Y.}~\bibnamefont{{Akiba}}},
  \bibinfo{author}{\bibfnamefont{A.~E.} \bibnamefont{{Anthony}}},
  \bibinfo{author}{\bibfnamefont{K.}~\bibnamefont{{Arnold}}},
  \bibinfo{author}{\bibfnamefont{M.}~\bibnamefont{{Atlas}}},
  \bibinfo{author}{\bibfnamefont{D.}~\bibnamefont{{Barron}}},
  \bibinfo{author}{\bibfnamefont{D.}~\bibnamefont{{Boettger}}},
  \bibinfo{author}{\bibfnamefont{J.}~\bibnamefont{{Borrill}}},
  \bibinfo{author}{\bibfnamefont{C.}~\bibnamefont{{Borys}}},
  \bibinfo{author}{\bibfnamefont{S.}~\bibnamefont{{Chapman}}},
  \bibnamefont{et~al.}, \bibinfo{journal}{Physical Review Letters}
  \textbf{\bibinfo{volume}{112}}, \bibinfo{eid}{131302} (\bibinfo{year}{2014}),
  \eprint{1312.6645}.

\bibitem[{\citenamefont{{Ade} et~al.}(2013)\citenamefont{{Ade}, {Akiba},
  {Anthony}, {Arnold}, {Barron}, {Boettger}, {Borrill}, {Chapman}, {Chinone},
  {Dobbs} et~al.}}]{pbear-eeeb/2013}
\bibinfo{author}{\bibfnamefont{P.~A.~R.} \bibnamefont{{Ade}}},
  \bibinfo{author}{\bibfnamefont{Y.}~\bibnamefont{{Akiba}}},
  \bibinfo{author}{\bibfnamefont{A.~E.} \bibnamefont{{Anthony}}},
  \bibinfo{author}{\bibfnamefont{K.}~\bibnamefont{{Arnold}}},
  \bibinfo{author}{\bibfnamefont{D.}~\bibnamefont{{Barron}}},
  \bibinfo{author}{\bibfnamefont{D.}~\bibnamefont{{Boettger}}},
  \bibinfo{author}{\bibfnamefont{J.}~\bibnamefont{{Borrill}}},
  \bibinfo{author}{\bibfnamefont{S.}~\bibnamefont{{Chapman}}},
  \bibinfo{author}{\bibfnamefont{Y.}~\bibnamefont{{Chinone}}},
  \bibinfo{author}{\bibfnamefont{M.}~\bibnamefont{{Dobbs}}},
  \bibnamefont{et~al.} (\bibinfo{collaboration}{{\textsc{Polarbear}}})
  (\bibinfo{year}{2013}), \eprint{1312.6646}.

\bibitem[{\citenamefont{{Hand} et~al.}(2013)\citenamefont{{Hand}, {Leauthaud},
  {Das}, {Sherwin}, {Addison}, {Bond}, {Calabrese}, {Charbonnier}, {Devlin},
  {Dunkley} et~al.}}]{hand/etal/2013}
\bibinfo{author}{\bibfnamefont{N.}~\bibnamefont{{Hand}}},
  \bibinfo{author}{\bibfnamefont{A.}~\bibnamefont{{Leauthaud}}},
  \bibinfo{author}{\bibfnamefont{S.}~\bibnamefont{{Das}}},
  \bibinfo{author}{\bibfnamefont{B.~D.} \bibnamefont{{Sherwin}}},
  \bibinfo{author}{\bibfnamefont{G.~E.} \bibnamefont{{Addison}}},
  \bibinfo{author}{\bibfnamefont{J.~R.} \bibnamefont{{Bond}}},
  \bibinfo{author}{\bibfnamefont{E.}~\bibnamefont{{Calabrese}}},
  \bibinfo{author}{\bibfnamefont{A.}~\bibnamefont{{Charbonnier}}},
  \bibinfo{author}{\bibfnamefont{M.~J.} \bibnamefont{{Devlin}}},
  \bibinfo{author}{\bibfnamefont{J.}~\bibnamefont{{Dunkley}}},
  \bibnamefont{et~al.} (\bibinfo{year}{2013}), \eprint{1311.6200}.

\bibitem[{\citenamefont{{Bianchini} et~al.}(2014)\citenamefont{{Bianchini},
  {Bielewicz}, {Lapi}, {Gonzalez-Nuevo}, {Baccigalupi}, {de Zotti}, {Danese},
  {Bourne}, {Cooray}, {Dunne} et~al.}}]{bianchini/etal/2014}
\bibinfo{author}{\bibfnamefont{F.}~\bibnamefont{{Bianchini}}},
  \bibinfo{author}{\bibfnamefont{P.}~\bibnamefont{{Bielewicz}}},
  \bibinfo{author}{\bibfnamefont{A.}~\bibnamefont{{Lapi}}},
  \bibinfo{author}{\bibfnamefont{J.}~\bibnamefont{{Gonzalez-Nuevo}}},
  \bibinfo{author}{\bibfnamefont{C.}~\bibnamefont{{Baccigalupi}}},
  \bibinfo{author}{\bibfnamefont{G.}~\bibnamefont{{de Zotti}}},
  \bibinfo{author}{\bibfnamefont{L.}~\bibnamefont{{Danese}}},
  \bibinfo{author}{\bibfnamefont{N.}~\bibnamefont{{Bourne}}},
  \bibinfo{author}{\bibfnamefont{A.}~\bibnamefont{{Cooray}}},
  \bibinfo{author}{\bibfnamefont{L.}~\bibnamefont{{Dunne}}},
  \bibnamefont{et~al.}, \bibinfo{journal}{ArXiv e-prints}
  (\bibinfo{year}{2014}), \eprint{1410.4502}.

\bibitem[{\citenamefont{{DiPompeo} et~al.}(2014)\citenamefont{{DiPompeo},
  {Myers}, {Hickox}, {Geach}, {Holder}, {Hainline}, and
  {Hall}}}]{dipompeo/etal/2014}
\bibinfo{author}{\bibfnamefont{M.~A.} \bibnamefont{{DiPompeo}}},
  \bibinfo{author}{\bibfnamefont{A.~D.} \bibnamefont{{Myers}}},
  \bibinfo{author}{\bibfnamefont{R.~C.} \bibnamefont{{Hickox}}},
  \bibinfo{author}{\bibfnamefont{J.~E.} \bibnamefont{{Geach}}},
  \bibinfo{author}{\bibfnamefont{G.}~\bibnamefont{{Holder}}},
  \bibinfo{author}{\bibfnamefont{K.~N.} \bibnamefont{{Hainline}}},
  \bibnamefont{and} \bibinfo{author}{\bibfnamefont{S.~W.}
  \bibnamefont{{Hall}}}, \bibinfo{journal}{ArXiv e-prints}
  (\bibinfo{year}{2014}), \eprint{1411.0527}.

\bibitem[{\citenamefont{{Fornengo} et~al.}(2014)\citenamefont{{Fornengo},
  {Perotto}, {Regis}, and {Camera}}}]{fornego/etal/2014}
\bibinfo{author}{\bibfnamefont{N.}~\bibnamefont{{Fornengo}}},
  \bibinfo{author}{\bibfnamefont{L.}~\bibnamefont{{Perotto}}},
  \bibinfo{author}{\bibfnamefont{M.}~\bibnamefont{{Regis}}}, \bibnamefont{and}
  \bibinfo{author}{\bibfnamefont{S.}~\bibnamefont{{Camera}}},
  \bibinfo{journal}{ArXiv e-prints}  (\bibinfo{year}{2014}),
  \eprint{1410.4997}.

\bibitem[{\citenamefont{{Hill} and {Spergel}}(2014)}]{HillSpergelLensing2014}
\bibinfo{author}{\bibfnamefont{J.~C.} \bibnamefont{{Hill}}} \bibnamefont{and}
  \bibinfo{author}{\bibfnamefont{D.~N.} \bibnamefont{{Spergel}}},
  \bibinfo{journal}{\jcap} \textbf{\bibinfo{volume}{2}}, \bibinfo{eid}{030}
  (\bibinfo{year}{2014}), \eprint{1312.4525}.

\bibitem[{\citenamefont{{Planck
  Collaboration}}(2014{\natexlab{c}})}]{planck_params/2013}
\bibinfo{author}{\bibnamefont{{Planck Collaboration}}}, \bibinfo{journal}{\aap}
  \textbf{\bibinfo{volume}{571}}, \bibinfo{eid}{A16}
  (\bibinfo{year}{2014}{\natexlab{c}}), \eprint{1303.5076}.

\bibitem[{\citenamefont{{Abazajian}
  et~al.}(2013{\natexlab{a}})\citenamefont{{Abazajian}, {Arnold}, {Austermann},
  {Benson}, {Bischoff}, {Bock}, {Bond}, {Borrill}, {Calabrese}, {Carlstrom}
  et~al.}}]{CMBS4Nus}
\bibinfo{author}{\bibfnamefont{K.~N.} \bibnamefont{{Abazajian}}},
  \bibinfo{author}{\bibfnamefont{K.}~\bibnamefont{{Arnold}}},
  \bibinfo{author}{\bibfnamefont{J.}~\bibnamefont{{Austermann}}},
  \bibinfo{author}{\bibfnamefont{B.~A.} \bibnamefont{{Benson}}},
  \bibinfo{author}{\bibfnamefont{C.}~\bibnamefont{{Bischoff}}},
  \bibinfo{author}{\bibfnamefont{J.}~\bibnamefont{{Bock}}},
  \bibinfo{author}{\bibfnamefont{J.~R.} \bibnamefont{{Bond}}},
  \bibinfo{author}{\bibfnamefont{J.}~\bibnamefont{{Borrill}}},
  \bibinfo{author}{\bibfnamefont{E.}~\bibnamefont{{Calabrese}}},
  \bibinfo{author}{\bibfnamefont{J.~E.} \bibnamefont{{Carlstrom}}},
  \bibnamefont{et~al.}, \bibinfo{journal}{ArXiv e-prints}
  (\bibinfo{year}{2013}{\natexlab{a}}), \eprint{1309.5383}.

\bibitem[{\citenamefont{{Abazajian}
  et~al.}(2013{\natexlab{b}})\citenamefont{{Abazajian}, {Arnold}, {Austermann},
  {Benson}, {Bischoff}, {Bock}, {Bond}, {Borrill}, {Buder}, {Burke}
  et~al.}}]{SnowmassInf2013}
\bibinfo{author}{\bibfnamefont{K.~N.} \bibnamefont{{Abazajian}}},
  \bibinfo{author}{\bibfnamefont{K.}~\bibnamefont{{Arnold}}},
  \bibinfo{author}{\bibfnamefont{J.}~\bibnamefont{{Austermann}}},
  \bibinfo{author}{\bibfnamefont{B.~A.} \bibnamefont{{Benson}}},
  \bibinfo{author}{\bibfnamefont{C.}~\bibnamefont{{Bischoff}}},
  \bibinfo{author}{\bibfnamefont{J.}~\bibnamefont{{Bock}}},
  \bibinfo{author}{\bibfnamefont{J.~R.} \bibnamefont{{Bond}}},
  \bibinfo{author}{\bibfnamefont{J.}~\bibnamefont{{Borrill}}},
  \bibinfo{author}{\bibfnamefont{I.}~\bibnamefont{{Buder}}},
  \bibinfo{author}{\bibfnamefont{D.~L.} \bibnamefont{{Burke}}},
  \bibnamefont{et~al.}, \bibinfo{journal}{ArXiv e-prints}
  (\bibinfo{year}{2013}{\natexlab{b}}), \eprint{1309.5381}.

\bibitem[{\citenamefont{Calabrese et~al.}(2014)\citenamefont{Calabrese,
  Hlo{\v{z}}ek, Battaglia, Bond, de~Bernardis, Devlin, Hajian, Henderson, Hil,
  Kosowsky et~al.}}]{calabrese2014precision}
\bibinfo{author}{\bibfnamefont{E.}~\bibnamefont{Calabrese}},
  \bibinfo{author}{\bibfnamefont{R.}~\bibnamefont{Hlo{\v{z}}ek}},
  \bibinfo{author}{\bibfnamefont{N.}~\bibnamefont{Battaglia}},
  \bibinfo{author}{\bibfnamefont{J.~R.} \bibnamefont{Bond}},
  \bibinfo{author}{\bibfnamefont{F.}~\bibnamefont{de~Bernardis}},
  \bibinfo{author}{\bibfnamefont{M.~J.} \bibnamefont{Devlin}},
  \bibinfo{author}{\bibfnamefont{A.}~\bibnamefont{Hajian}},
  \bibinfo{author}{\bibfnamefont{S.}~\bibnamefont{Henderson}},
  \bibinfo{author}{\bibfnamefont{J.~C.} \bibnamefont{Hil}},
  \bibinfo{author}{\bibfnamefont{A.}~\bibnamefont{Kosowsky}},
  \bibnamefont{et~al.}, \bibinfo{journal}{Journal of Cosmology and
  Astroparticle Physics} \textbf{\bibinfo{volume}{2014}}, \bibinfo{pages}{010}
  (\bibinfo{year}{2014}).

\bibitem[{\citenamefont{Seljak and Zaldarriaga}(2000)}]{seljak2000lensing}
\bibinfo{author}{\bibfnamefont{U.}~\bibnamefont{Seljak}} \bibnamefont{and}
  \bibinfo{author}{\bibfnamefont{M.}~\bibnamefont{Zaldarriaga}},
  \bibinfo{journal}{The Astrophysical Journal} \textbf{\bibinfo{volume}{538}},
  \bibinfo{pages}{57} (\bibinfo{year}{2000}).

\bibitem[{\citenamefont{Zaldarriaga}(2000)}]{zaldarriaga2000lensing}
\bibinfo{author}{\bibfnamefont{M.}~\bibnamefont{Zaldarriaga}},
  \bibinfo{journal}{Physical Review D} \textbf{\bibinfo{volume}{62}},
  \bibinfo{pages}{063510} (\bibinfo{year}{2000}).

\bibitem[{\citenamefont{{Dodelson} and {Starkman}}(2003)}]{dodelsonGalaxy2003}
\bibinfo{author}{\bibfnamefont{S.}~\bibnamefont{{Dodelson}}} \bibnamefont{and}
  \bibinfo{author}{\bibfnamefont{G.~D.} \bibnamefont{{Starkman}}},
  \bibinfo{journal}{ArXiv Astrophysics e-prints}  (\bibinfo{year}{2003}),
  \eprint{astro-ph/0305467}.

\bibitem[{\citenamefont{Holder and Kosowsky}(2004)}]{holder2004gravitational}
\bibinfo{author}{\bibfnamefont{G.}~\bibnamefont{Holder}} \bibnamefont{and}
  \bibinfo{author}{\bibfnamefont{A.}~\bibnamefont{Kosowsky}},
  \bibinfo{journal}{The Astrophysical Journal} \textbf{\bibinfo{volume}{616}},
  \bibinfo{pages}{8} (\bibinfo{year}{2004}).

\bibitem[{\citenamefont{Dodelson}(2004)}]{dodelson2004cmb}
\bibinfo{author}{\bibfnamefont{S.}~\bibnamefont{Dodelson}},
  \bibinfo{journal}{Physical Review D} \textbf{\bibinfo{volume}{70}},
  \bibinfo{pages}{023009} (\bibinfo{year}{2004}).

\bibitem[{\citenamefont{{Vale} et~al.}(2004)\citenamefont{{Vale}, {Amblard},
  and {White}}}]{ValeAmblardWhite2004}
\bibinfo{author}{\bibfnamefont{C.}~\bibnamefont{{Vale}}},
  \bibinfo{author}{\bibfnamefont{A.}~\bibnamefont{{Amblard}}},
  \bibnamefont{and} \bibinfo{author}{\bibfnamefont{M.}~\bibnamefont{{White}}},
  \bibinfo{journal}{\na} \textbf{\bibinfo{volume}{10}}, \bibinfo{pages}{1}
  (\bibinfo{year}{2004}), \eprint{astro-ph/0402004}.

\bibitem[{\citenamefont{{Maturi} et~al.}(2005)\citenamefont{{Maturi},
  {Bartelmann}, {Meneghetti}, and {Moscardini}}}]{maturi.bartelmann.ea:2005}
\bibinfo{author}{\bibfnamefont{M.}~\bibnamefont{{Maturi}}},
  \bibinfo{author}{\bibfnamefont{M.}~\bibnamefont{{Bartelmann}}},
  \bibinfo{author}{\bibfnamefont{M.}~\bibnamefont{{Meneghetti}}},
  \bibnamefont{and}
  \bibinfo{author}{\bibfnamefont{L.}~\bibnamefont{{Moscardini}}},
  \bibinfo{journal}{\aap} \textbf{\bibinfo{volume}{436}}, \bibinfo{pages}{37}
  (\bibinfo{year}{2005}), \eprint{arXiv:astro-ph/0408064}.

\bibitem[{\citenamefont{{Lewis} and {King}}(2006)}]{LewisKing2006}
\bibinfo{author}{\bibfnamefont{A.}~\bibnamefont{{Lewis}}} \bibnamefont{and}
  \bibinfo{author}{\bibfnamefont{L.}~\bibnamefont{{King}}},
  \bibinfo{journal}{\prd} \textbf{\bibinfo{volume}{73}}, \bibinfo{eid}{063006}
  (\bibinfo{year}{2006}), \eprint{astro-ph/0512104}.

\bibitem[{\citenamefont{{Hu} et~al.}(2007)\citenamefont{{Hu}, {Holz}, and
  {Vale}}}]{HuHolzVale2007}
\bibinfo{author}{\bibfnamefont{W.}~\bibnamefont{{Hu}}},
  \bibinfo{author}{\bibfnamefont{D.~E.} \bibnamefont{{Holz}}},
  \bibnamefont{and} \bibinfo{author}{\bibfnamefont{C.}~\bibnamefont{{Vale}}},
  \bibinfo{journal}{\prd} \textbf{\bibinfo{volume}{76}}, \bibinfo{eid}{127301}
  (\bibinfo{year}{2007}), \eprint{0708.4391}.

\bibitem[{\citenamefont{Hu et~al.}(2007)\citenamefont{Hu, DeDeo, and
  Vale}}]{hu2007cluster}
\bibinfo{author}{\bibfnamefont{W.}~\bibnamefont{Hu}},
  \bibinfo{author}{\bibfnamefont{S.}~\bibnamefont{DeDeo}}, \bibnamefont{and}
  \bibinfo{author}{\bibfnamefont{C.}~\bibnamefont{Vale}}, \bibinfo{journal}{New
  Journal of Physics} \textbf{\bibinfo{volume}{9}}, \bibinfo{pages}{441}
  (\bibinfo{year}{2007}).

\bibitem[{\citenamefont{{Yoo} and {Zaldarriaga}}(2008)}]{Yoo2008}
\bibinfo{author}{\bibfnamefont{J.}~\bibnamefont{{Yoo}}} \bibnamefont{and}
  \bibinfo{author}{\bibfnamefont{M.}~\bibnamefont{{Zaldarriaga}}},
  \bibinfo{journal}{\prd} \textbf{\bibinfo{volume}{78}}, \bibinfo{eid}{083002}
  (\bibinfo{year}{2008}), \eprint{0805.2155}.

\bibitem[{\citenamefont{{Yoo} et~al.}(2010)\citenamefont{{Yoo}, {Zaldarriaga},
  and {Hernquist}}}]{Yoo2010}
\bibinfo{author}{\bibfnamefont{J.}~\bibnamefont{{Yoo}}},
  \bibinfo{author}{\bibfnamefont{M.}~\bibnamefont{{Zaldarriaga}}},
  \bibnamefont{and}
  \bibinfo{author}{\bibfnamefont{L.}~\bibnamefont{{Hernquist}}},
  \bibinfo{journal}{\prd} \textbf{\bibinfo{volume}{81}}, \bibinfo{eid}{123006}
  (\bibinfo{year}{2010}), \eprint{1005.0847}.

\bibitem[{\citenamefont{{Melin} and {Bartlett}}(2014)}]{Melin2014}
\bibinfo{author}{\bibfnamefont{J.-B.} \bibnamefont{{Melin}}} \bibnamefont{and}
  \bibinfo{author}{\bibfnamefont{J.~G.} \bibnamefont{{Bartlett}}},
  \bibinfo{journal}{ArXiv e-prints}  (\bibinfo{year}{2014}),
  \eprint{1408.5633}.

\bibitem[{\citenamefont{Eisenstein et~al.}(2011)\citenamefont{Eisenstein,
  Weinberg, Agol, Aihara, Prieto, Anderson, Arns, Aubourg, Bailey, Balbinot
  et~al.}}]{eisenstein2011sdss}
\bibinfo{author}{\bibfnamefont{D.~J.} \bibnamefont{Eisenstein}},
  \bibinfo{author}{\bibfnamefont{D.~H.} \bibnamefont{Weinberg}},
  \bibinfo{author}{\bibfnamefont{E.}~\bibnamefont{Agol}},
  \bibinfo{author}{\bibfnamefont{H.}~\bibnamefont{Aihara}},
  \bibinfo{author}{\bibfnamefont{C.~A.} \bibnamefont{Prieto}},
  \bibinfo{author}{\bibfnamefont{S.~F.} \bibnamefont{Anderson}},
  \bibinfo{author}{\bibfnamefont{J.~A.} \bibnamefont{Arns}},
  \bibinfo{author}{\bibfnamefont{{\'E}.}~\bibnamefont{Aubourg}},
  \bibinfo{author}{\bibfnamefont{S.}~\bibnamefont{Bailey}},
  \bibinfo{author}{\bibfnamefont{E.}~\bibnamefont{Balbinot}},
  \bibnamefont{et~al.}, \bibinfo{journal}{The Astronomical Journal}
  \textbf{\bibinfo{volume}{142}}, \bibinfo{pages}{72} (\bibinfo{year}{2011}).

\bibitem[{\citenamefont{Dawson et~al.}(2013)\citenamefont{Dawson, Schlegel,
  Ahn, Anderson, Aubourg, Bailey, Barkhouser, Bautista, Beifiori, Berlind
  et~al.}}]{dawson2013baryon}
\bibinfo{author}{\bibfnamefont{K.~S.} \bibnamefont{Dawson}},
  \bibinfo{author}{\bibfnamefont{D.~J.} \bibnamefont{Schlegel}},
  \bibinfo{author}{\bibfnamefont{C.~P.} \bibnamefont{Ahn}},
  \bibinfo{author}{\bibfnamefont{S.~F.} \bibnamefont{Anderson}},
  \bibinfo{author}{\bibfnamefont{{\'E}.}~\bibnamefont{Aubourg}},
  \bibinfo{author}{\bibfnamefont{S.}~\bibnamefont{Bailey}},
  \bibinfo{author}{\bibfnamefont{R.~H.} \bibnamefont{Barkhouser}},
  \bibinfo{author}{\bibfnamefont{J.~E.} \bibnamefont{Bautista}},
  \bibinfo{author}{\bibfnamefont{A.}~\bibnamefont{Beifiori}},
  \bibinfo{author}{\bibfnamefont{A.~A.} \bibnamefont{Berlind}},
  \bibnamefont{et~al.}, \bibinfo{journal}{The Astronomical Journal}
  \textbf{\bibinfo{volume}{145}}, \bibinfo{pages}{10} (\bibinfo{year}{2013}).

\bibitem[{\citenamefont{Ahn et~al.}(2014)\citenamefont{Ahn, Alexandroff,
  Allende~Prieto, Anders, Anderson, Anderton, Andrews, Aubourg, Bailey, Bastien
  et~al.}}]{ahn2014tenth}
\bibinfo{author}{\bibfnamefont{C.~P.} \bibnamefont{Ahn}},
  \bibinfo{author}{\bibfnamefont{R.}~\bibnamefont{Alexandroff}},
  \bibinfo{author}{\bibfnamefont{C.}~\bibnamefont{Allende~Prieto}},
  \bibinfo{author}{\bibfnamefont{F.}~\bibnamefont{Anders}},
  \bibinfo{author}{\bibfnamefont{S.~F.} \bibnamefont{Anderson}},
  \bibinfo{author}{\bibfnamefont{T.}~\bibnamefont{Anderton}},
  \bibinfo{author}{\bibfnamefont{B.~H.} \bibnamefont{Andrews}},
  \bibinfo{author}{\bibfnamefont{{\'E}.}~\bibnamefont{Aubourg}},
  \bibinfo{author}{\bibfnamefont{S.}~\bibnamefont{Bailey}},
  \bibinfo{author}{\bibfnamefont{F.~A.} \bibnamefont{Bastien}},
  \bibnamefont{et~al.}, \bibinfo{journal}{The Astrophysical Journal Supplement
  Series} \textbf{\bibinfo{volume}{211}}, \bibinfo{pages}{17}
  (\bibinfo{year}{2014}).

\bibitem[{\citenamefont{{Niemack} et~al.}(2010)\citenamefont{{Niemack}, {Ade},
  {Aguirre}, {Barrientos}, {Beall}, {Bond}, {Britton}, {Cho}, {Das}, {Devlin}
  et~al.}}]{niemack/etal/2010}
\bibinfo{author}{\bibfnamefont{M.~D.} \bibnamefont{{Niemack}}},
  \bibinfo{author}{\bibfnamefont{P.~A.~R.} \bibnamefont{{Ade}}},
  \bibinfo{author}{\bibfnamefont{J.}~\bibnamefont{{Aguirre}}},
  \bibinfo{author}{\bibfnamefont{F.}~\bibnamefont{{Barrientos}}},
  \bibinfo{author}{\bibfnamefont{J.~A.} \bibnamefont{{Beall}}},
  \bibinfo{author}{\bibfnamefont{J.~R.} \bibnamefont{{Bond}}},
  \bibinfo{author}{\bibfnamefont{J.}~\bibnamefont{{Britton}}},
  \bibinfo{author}{\bibfnamefont{H.~M.} \bibnamefont{{Cho}}},
  \bibinfo{author}{\bibfnamefont{S.}~\bibnamefont{{Das}}},
  \bibinfo{author}{\bibfnamefont{M.~J.} \bibnamefont{{Devlin}}},
  \bibnamefont{et~al.}, in \emph{\bibinfo{booktitle}{Society of Photo-Optical
  Instrumentation Engineers (SPIE) Conference Series}} (\bibinfo{year}{2010}),
  vol. \bibinfo{volume}{7741}, \eprint{1006.5049}.

\bibitem[{\citenamefont{{Naess} et~al.}(2014)\citenamefont{{Naess},
  {Hasselfield}, {McMahon}, {Niemack}, {Addison}, {Ade}, {Allison}, {Amiri},
  {Battaglia}, {Beall} et~al.}}]{Naess2014}
\bibinfo{author}{\bibfnamefont{S.}~\bibnamefont{{Naess}}},
  \bibinfo{author}{\bibfnamefont{M.}~\bibnamefont{{Hasselfield}}},
  \bibinfo{author}{\bibfnamefont{J.}~\bibnamefont{{McMahon}}},
  \bibinfo{author}{\bibfnamefont{M.~D.} \bibnamefont{{Niemack}}},
  \bibinfo{author}{\bibfnamefont{G.~E.} \bibnamefont{{Addison}}},
  \bibinfo{author}{\bibfnamefont{P.~A.~R.} \bibnamefont{{Ade}}},
  \bibinfo{author}{\bibfnamefont{R.}~\bibnamefont{{Allison}}},
  \bibinfo{author}{\bibfnamefont{M.}~\bibnamefont{{Amiri}}},
  \bibinfo{author}{\bibfnamefont{N.}~\bibnamefont{{Battaglia}}},
  \bibinfo{author}{\bibfnamefont{J.~A.} \bibnamefont{{Beall}}},
  \bibnamefont{et~al.}, \bibinfo{journal}{\jcap} \textbf{\bibinfo{volume}{10}},
  \bibinfo{eid}{007} (\bibinfo{year}{2014}), \eprint{1405.5524}.

\bibitem[{\citenamefont{{Planck Collaboration}}(2013)}]{planck_mission/2013}
\bibinfo{author}{\bibnamefont{{Planck Collaboration}}}
  (\bibinfo{collaboration}{{Planck}}) (\bibinfo{year}{2013}),
  \eprint{1303.5062}.

\bibitem[{\citenamefont{{Louis} et~al.}(2014)\citenamefont{{Louis}, {Addison},
  {Hasselfield}, {Bond}, {Calabrese}, {Das}, {Devlin}, {Dunkley}, {D{\"u}nner},
  {Gralla} et~al.}}]{louis/etal/2014}
\bibinfo{author}{\bibfnamefont{T.}~\bibnamefont{{Louis}}},
  \bibinfo{author}{\bibfnamefont{G.~E.} \bibnamefont{{Addison}}},
  \bibinfo{author}{\bibfnamefont{M.}~\bibnamefont{{Hasselfield}}},
  \bibinfo{author}{\bibfnamefont{J.~R.} \bibnamefont{{Bond}}},
  \bibinfo{author}{\bibfnamefont{E.}~\bibnamefont{{Calabrese}}},
  \bibinfo{author}{\bibfnamefont{S.}~\bibnamefont{{Das}}},
  \bibinfo{author}{\bibfnamefont{M.~J.} \bibnamefont{{Devlin}}},
  \bibinfo{author}{\bibfnamefont{J.}~\bibnamefont{{Dunkley}}},
  \bibinfo{author}{\bibfnamefont{R.}~\bibnamefont{{D{\"u}nner}}},
  \bibinfo{author}{\bibfnamefont{M.}~\bibnamefont{{Gralla}}},
  \bibnamefont{et~al.}, \bibinfo{journal}{\jcap} \textbf{\bibinfo{volume}{7}},
  \bibinfo{eid}{016} (\bibinfo{year}{2014}), \eprint{1403.0608}.

\bibitem[{\citenamefont{York et~al.}(2000)\citenamefont{York, Adelman,
  Anderson~Jr, Anderson, Annis, Bahcall, Bakken, Barkhouser, Bastian, Berman
  et~al.}}]{york2000sloan}
\bibinfo{author}{\bibfnamefont{D.~G.} \bibnamefont{York}},
  \bibinfo{author}{\bibfnamefont{J.}~\bibnamefont{Adelman}},
  \bibinfo{author}{\bibfnamefont{J.~E.} \bibnamefont{Anderson~Jr}},
  \bibinfo{author}{\bibfnamefont{S.~F.} \bibnamefont{Anderson}},
  \bibinfo{author}{\bibfnamefont{J.}~\bibnamefont{Annis}},
  \bibinfo{author}{\bibfnamefont{N.~A.} \bibnamefont{Bahcall}},
  \bibinfo{author}{\bibfnamefont{J.}~\bibnamefont{Bakken}},
  \bibinfo{author}{\bibfnamefont{R.}~\bibnamefont{Barkhouser}},
  \bibinfo{author}{\bibfnamefont{S.}~\bibnamefont{Bastian}},
  \bibinfo{author}{\bibfnamefont{E.}~\bibnamefont{Berman}},
  \bibnamefont{et~al.}, \bibinfo{journal}{The Astronomical Journal}
  \textbf{\bibinfo{volume}{120}}, \bibinfo{pages}{1579} (\bibinfo{year}{2000}).

\bibitem[{\citenamefont{Gunn et~al.}(2006)\citenamefont{Gunn, Siegmund,
  Mannery, Owen, Hull, Leger, Carey, Knapp, York, Boroski et~al.}}]{gunn20062}
\bibinfo{author}{\bibfnamefont{J.~E.} \bibnamefont{Gunn}},
  \bibinfo{author}{\bibfnamefont{W.~A.} \bibnamefont{Siegmund}},
  \bibinfo{author}{\bibfnamefont{E.~J.} \bibnamefont{Mannery}},
  \bibinfo{author}{\bibfnamefont{R.~E.} \bibnamefont{Owen}},
  \bibinfo{author}{\bibfnamefont{C.~L.} \bibnamefont{Hull}},
  \bibinfo{author}{\bibfnamefont{R.~F.} \bibnamefont{Leger}},
  \bibinfo{author}{\bibfnamefont{L.~N.} \bibnamefont{Carey}},
  \bibinfo{author}{\bibfnamefont{G.~R.} \bibnamefont{Knapp}},
  \bibinfo{author}{\bibfnamefont{D.~G.} \bibnamefont{York}},
  \bibinfo{author}{\bibfnamefont{W.~N.} \bibnamefont{Boroski}},
  \bibnamefont{et~al.}, \bibinfo{journal}{The Astronomical Journal}
  \textbf{\bibinfo{volume}{131}}, \bibinfo{pages}{2332} (\bibinfo{year}{2006}).

\bibitem[{\citenamefont{{Eisenstein} et~al.}(2001)\citenamefont{{Eisenstein},
  {Annis}, {Gunn}, {Szalay}, {Connolly}, {Nichol}, {Bahcall}, {Bernardi},
  {Burles}, {Castander} et~al.}}]{Eisenstein2001}
\bibinfo{author}{\bibfnamefont{D.~J.} \bibnamefont{{Eisenstein}}},
  \bibinfo{author}{\bibfnamefont{J.}~\bibnamefont{{Annis}}},
  \bibinfo{author}{\bibfnamefont{J.~E.} \bibnamefont{{Gunn}}},
  \bibinfo{author}{\bibfnamefont{A.~S.} \bibnamefont{{Szalay}}},
  \bibinfo{author}{\bibfnamefont{A.~J.} \bibnamefont{{Connolly}}},
  \bibinfo{author}{\bibfnamefont{R.~C.} \bibnamefont{{Nichol}}},
  \bibinfo{author}{\bibfnamefont{N.~A.} \bibnamefont{{Bahcall}}},
  \bibinfo{author}{\bibfnamefont{M.}~\bibnamefont{{Bernardi}}},
  \bibinfo{author}{\bibfnamefont{S.}~\bibnamefont{{Burles}}},
  \bibinfo{author}{\bibfnamefont{F.~J.} \bibnamefont{{Castander}}},
  \bibnamefont{et~al.}, \bibinfo{journal}{\aj} \textbf{\bibinfo{volume}{122}},
  \bibinfo{pages}{2267} (\bibinfo{year}{2001}), \eprint{astro-ph/0108153}.

\bibitem[{\citenamefont{{Anderson} et~al.}(2012)\citenamefont{{Anderson},
  {Aubourg}, {Bailey}, {Bizyaev}, {Blanton}, {Bolton}, {Brinkmann},
  {Brownstein}, {Burden}, {Cuesta} et~al.}}]{andersonCMASS2012}
\bibinfo{author}{\bibfnamefont{L.}~\bibnamefont{{Anderson}}},
  \bibinfo{author}{\bibfnamefont{E.}~\bibnamefont{{Aubourg}}},
  \bibinfo{author}{\bibfnamefont{S.}~\bibnamefont{{Bailey}}},
  \bibinfo{author}{\bibfnamefont{D.}~\bibnamefont{{Bizyaev}}},
  \bibinfo{author}{\bibfnamefont{M.}~\bibnamefont{{Blanton}}},
  \bibinfo{author}{\bibfnamefont{A.~S.} \bibnamefont{{Bolton}}},
  \bibinfo{author}{\bibfnamefont{J.}~\bibnamefont{{Brinkmann}}},
  \bibinfo{author}{\bibfnamefont{J.~R.} \bibnamefont{{Brownstein}}},
  \bibinfo{author}{\bibfnamefont{A.}~\bibnamefont{{Burden}}},
  \bibinfo{author}{\bibfnamefont{A.~J.} \bibnamefont{{Cuesta}}},
  \bibnamefont{et~al.}, \bibinfo{journal}{\mnras}
  \textbf{\bibinfo{volume}{427}}, \bibinfo{pages}{3435} (\bibinfo{year}{2012}),
  \eprint{1203.6594}.

\bibitem[{\citenamefont{{Reid} et~al.}(2012)\citenamefont{{Reid}, {Samushia},
  {White}, {Percival}, {Manera}, {Padmanabhan}, {Ross}, {S{\'a}nchez},
  {Bailey}, {Bizyaev} et~al.}}]{reidCMASS2012}
\bibinfo{author}{\bibfnamefont{B.~A.} \bibnamefont{{Reid}}},
  \bibinfo{author}{\bibfnamefont{L.}~\bibnamefont{{Samushia}}},
  \bibinfo{author}{\bibfnamefont{M.}~\bibnamefont{{White}}},
  \bibinfo{author}{\bibfnamefont{W.~J.} \bibnamefont{{Percival}}},
  \bibinfo{author}{\bibfnamefont{M.}~\bibnamefont{{Manera}}},
  \bibinfo{author}{\bibfnamefont{N.}~\bibnamefont{{Padmanabhan}}},
  \bibinfo{author}{\bibfnamefont{A.~J.} \bibnamefont{{Ross}}},
  \bibinfo{author}{\bibfnamefont{A.~G.} \bibnamefont{{S{\'a}nchez}}},
  \bibinfo{author}{\bibfnamefont{S.}~\bibnamefont{{Bailey}}},
  \bibinfo{author}{\bibfnamefont{D.}~\bibnamefont{{Bizyaev}}},
  \bibnamefont{et~al.}, \bibinfo{journal}{\mnras}
  \textbf{\bibinfo{volume}{426}}, \bibinfo{pages}{2719} (\bibinfo{year}{2012}),
  \eprint{1203.6641}.

\bibitem[{\citenamefont{Miyatake et~al.}(2013)\citenamefont{Miyatake, More,
  Mandelbaum, Takada, Spergel, Kneib, Schneider, Brinkmann, Brownstein
  et~al.}}]{miyatake2013weak}
\bibinfo{author}{\bibfnamefont{H.}~\bibnamefont{Miyatake}},
  \bibinfo{author}{\bibfnamefont{S.}~\bibnamefont{More}},
  \bibinfo{author}{\bibfnamefont{R.}~\bibnamefont{Mandelbaum}},
  \bibinfo{author}{\bibfnamefont{M.}~\bibnamefont{Takada}},
  \bibinfo{author}{\bibfnamefont{D.}~\bibnamefont{Spergel}},
  \bibinfo{author}{\bibfnamefont{J.-P.} \bibnamefont{Kneib}},
  \bibinfo{author}{\bibfnamefont{D.~P.} \bibnamefont{Schneider}},
  \bibinfo{author}{\bibfnamefont{J.}~\bibnamefont{Brinkmann}},
  \bibinfo{author}{\bibfnamefont{J.~R.} \bibnamefont{Brownstein}},
  \bibnamefont{et~al.}, \bibinfo{journal}{arXiv preprint arXiv:1311.1480}
  (\bibinfo{year}{2013}).

\bibitem[{\citenamefont{{Heymans} et~al.}(2012)\citenamefont{{Heymans}, {Van
  Waerbeke}, {Miller}, {Erben}, {Hildebrandt}, {Hoekstra}, {Kitching},
  {Mellier}, {Simon}, {Bonnett} et~al.}}]{cfhtlens2012}
\bibinfo{author}{\bibfnamefont{C.}~\bibnamefont{{Heymans}}},
  \bibinfo{author}{\bibfnamefont{L.}~\bibnamefont{{Van Waerbeke}}},
  \bibinfo{author}{\bibfnamefont{L.}~\bibnamefont{{Miller}}},
  \bibinfo{author}{\bibfnamefont{T.}~\bibnamefont{{Erben}}},
  \bibinfo{author}{\bibfnamefont{H.}~\bibnamefont{{Hildebrandt}}},
  \bibinfo{author}{\bibfnamefont{H.}~\bibnamefont{{Hoekstra}}},
  \bibinfo{author}{\bibfnamefont{T.~D.} \bibnamefont{{Kitching}}},
  \bibinfo{author}{\bibfnamefont{Y.}~\bibnamefont{{Mellier}}},
  \bibinfo{author}{\bibfnamefont{P.}~\bibnamefont{{Simon}}},
  \bibinfo{author}{\bibfnamefont{C.}~\bibnamefont{{Bonnett}}},
  \bibnamefont{et~al.}, \bibinfo{journal}{\mnras}
  \textbf{\bibinfo{volume}{427}}, \bibinfo{pages}{146} (\bibinfo{year}{2012}),
  \eprint{1210.0032}.

\bibitem[{\citenamefont{Hu and Okamoto}(2002)}]{hu2002mass}
\bibinfo{author}{\bibfnamefont{W.}~\bibnamefont{Hu}} \bibnamefont{and}
  \bibinfo{author}{\bibfnamefont{T.}~\bibnamefont{Okamoto}},
  \bibinfo{journal}{The Astrophysical Journal} \textbf{\bibinfo{volume}{574}},
  \bibinfo{pages}{566} (\bibinfo{year}{2002}).

\bibitem[{\citenamefont{{Hanson} et~al.}(2009)\citenamefont{{Hanson}, {Rocha},
  and {G{\'o}rski}}}]{hanson2009}
\bibinfo{author}{\bibfnamefont{D.}~\bibnamefont{{Hanson}}},
  \bibinfo{author}{\bibfnamefont{G.}~\bibnamefont{{Rocha}}}, \bibnamefont{and}
  \bibinfo{author}{\bibfnamefont{K.}~\bibnamefont{{G{\'o}rski}}},
  \bibinfo{journal}{\mnras} \textbf{\bibinfo{volume}{400}},
  \bibinfo{pages}{2169} (\bibinfo{year}{2009}), \eprint{0907.1927}.

\bibitem[{\citenamefont{{Namikawa} et~al.}(2013)\citenamefont{{Namikawa},
  {Hanson}, and {Takahashi}}}]{namikawa2013}
\bibinfo{author}{\bibfnamefont{T.}~\bibnamefont{{Namikawa}}},
  \bibinfo{author}{\bibfnamefont{D.}~\bibnamefont{{Hanson}}}, \bibnamefont{and}
  \bibinfo{author}{\bibfnamefont{R.}~\bibnamefont{{Takahashi}}},
  \bibinfo{journal}{\mnras} \textbf{\bibinfo{volume}{431}},
  \bibinfo{pages}{609} (\bibinfo{year}{2013}), \eprint{1209.0091}.

\bibitem[{\citenamefont{Navarro et~al.}(1997)\citenamefont{Navarro, Frenk, and
  White}}]{navarro1997universal}
\bibinfo{author}{\bibfnamefont{J.~F.} \bibnamefont{Navarro}},
  \bibinfo{author}{\bibfnamefont{C.~S.} \bibnamefont{Frenk}}, \bibnamefont{and}
  \bibinfo{author}{\bibfnamefont{S.~D.} \bibnamefont{White}},
  \bibinfo{journal}{The Astrophysical Journal} \textbf{\bibinfo{volume}{490}},
  \bibinfo{pages}{493} (\bibinfo{year}{1997}).

\bibitem[{\citenamefont{{Bartelmann}}(1996)}]{Bartlemann96}
\bibinfo{author}{\bibfnamefont{M.}~\bibnamefont{{Bartelmann}}},
  \bibinfo{journal}{\aap} \textbf{\bibinfo{volume}{313}}, \bibinfo{pages}{697}
  (\bibinfo{year}{1996}), \eprint{astro-ph/9602053}.

\bibitem[{\citenamefont{{Macci{\`o}} et~al.}(2007)\citenamefont{{Macci{\`o}},
  {Dutton}, {van den Bosch}, {Moore}, {Potter}, and {Stadel}}}]{Maccio2007}
\bibinfo{author}{\bibfnamefont{A.~V.} \bibnamefont{{Macci{\`o}}}},
  \bibinfo{author}{\bibfnamefont{A.~A.} \bibnamefont{{Dutton}}},
  \bibinfo{author}{\bibfnamefont{F.~C.} \bibnamefont{{van den Bosch}}},
  \bibinfo{author}{\bibfnamefont{B.}~\bibnamefont{{Moore}}},
  \bibinfo{author}{\bibfnamefont{D.}~\bibnamefont{{Potter}}}, \bibnamefont{and}
  \bibinfo{author}{\bibfnamefont{J.}~\bibnamefont{{Stadel}}},
  \bibinfo{journal}{\mnras} \textbf{\bibinfo{volume}{378}}, \bibinfo{pages}{55}
  (\bibinfo{year}{2007}), \eprint{astro-ph/0608157}.

\bibitem[{\citenamefont{Cooray et~al.}(2005)\citenamefont{Cooray, Kamionkowski,
  and Caldwell}}]{cooray2005}
\bibinfo{author}{\bibfnamefont{A.}~\bibnamefont{Cooray}},
  \bibinfo{author}{\bibfnamefont{M.}~\bibnamefont{Kamionkowski}},
  \bibnamefont{and} \bibinfo{author}{\bibfnamefont{R.~R.}
  \bibnamefont{Caldwell}}, \bibinfo{journal}{Physical Review D}
  \textbf{\bibinfo{volume}{71}}, \bibinfo{pages}{123527}
  (\bibinfo{year}{2005}).

\bibitem[{\citenamefont{Sherwin et~al.}(2012)\citenamefont{Sherwin, Das,
  Hajian, Addison, Bond, Crichton, Devlin, Dunkley, Gralla, Halpern
  et~al.}}]{sherwin2012}
\bibinfo{author}{\bibfnamefont{B.~D.} \bibnamefont{Sherwin}},
  \bibinfo{author}{\bibfnamefont{S.}~\bibnamefont{Das}},
  \bibinfo{author}{\bibfnamefont{A.}~\bibnamefont{Hajian}},
  \bibinfo{author}{\bibfnamefont{G.}~\bibnamefont{Addison}},
  \bibinfo{author}{\bibfnamefont{J.~R.} \bibnamefont{Bond}},
  \bibinfo{author}{\bibfnamefont{D.}~\bibnamefont{Crichton}},
  \bibinfo{author}{\bibfnamefont{M.~J.} \bibnamefont{Devlin}},
  \bibinfo{author}{\bibfnamefont{J.}~\bibnamefont{Dunkley}},
  \bibinfo{author}{\bibfnamefont{M.~B.} \bibnamefont{Gralla}},
  \bibinfo{author}{\bibfnamefont{M.}~\bibnamefont{Halpern}},
  \bibnamefont{et~al.}, \bibinfo{journal}{Physical Review D}
  \textbf{\bibinfo{volume}{86}}, \bibinfo{pages}{083006}
  (\bibinfo{year}{2012}).

\bibitem[{\citenamefont{{Sunyaev} and {Zel'dovich}}(1970)}]{SZ1970}
\bibinfo{author}{\bibfnamefont{R.~A.} \bibnamefont{{Sunyaev}}}
  \bibnamefont{and} \bibinfo{author}{\bibfnamefont{Y.~B.}
  \bibnamefont{{Zel'dovich}}}, \bibinfo{journal}{Comments on Astrophysics and
  Space Physics} \textbf{\bibinfo{volume}{2}}, \bibinfo{pages}{66}
  (\bibinfo{year}{1970}).

\bibitem[{\citenamefont{{Sunyaev} and {Zel'dovich}}(1972)}]{SZ1972}
\bibinfo{author}{\bibfnamefont{R.~A.} \bibnamefont{{Sunyaev}}}
  \bibnamefont{and} \bibinfo{author}{\bibfnamefont{Y.~B.}
  \bibnamefont{{Zel'dovich}}}, \bibinfo{journal}{Comments on Astrophysics and
  Space Physics} \textbf{\bibinfo{volume}{4}}, \bibinfo{pages}{173}
  (\bibinfo{year}{1972}).

\bibitem[{\citenamefont{{Hand} et~al.}(2011)\citenamefont{{Hand}, {Appel},
  {Battaglia}, {Bond}, {Das}, {Devlin}, {Dunkley}, {D{\"u}nner},
  {Essinger-Hileman}, {Fowler} et~al.}}]{HandSZ}
\bibinfo{author}{\bibfnamefont{N.}~\bibnamefont{{Hand}}},
  \bibinfo{author}{\bibfnamefont{J.~W.} \bibnamefont{{Appel}}},
  \bibinfo{author}{\bibfnamefont{N.}~\bibnamefont{{Battaglia}}},
  \bibinfo{author}{\bibfnamefont{J.~R.} \bibnamefont{{Bond}}},
  \bibinfo{author}{\bibfnamefont{S.}~\bibnamefont{{Das}}},
  \bibinfo{author}{\bibfnamefont{M.~J.} \bibnamefont{{Devlin}}},
  \bibinfo{author}{\bibfnamefont{J.}~\bibnamefont{{Dunkley}}},
  \bibinfo{author}{\bibfnamefont{R.}~\bibnamefont{{D{\"u}nner}}},
  \bibinfo{author}{\bibfnamefont{T.}~\bibnamefont{{Essinger-Hileman}}},
  \bibinfo{author}{\bibfnamefont{J.~W.} \bibnamefont{{Fowler}}},
  \bibnamefont{et~al.}, \bibinfo{journal}{\apj} \textbf{\bibinfo{volume}{736}},
  \bibinfo{eid}{39} (\bibinfo{year}{2011}), \eprint{1101.1951}.

\bibitem[{\citenamefont{{Ma} and {Fry}}(2000)}]{MaFry2Halo}
\bibinfo{author}{\bibfnamefont{C.-P.} \bibnamefont{{Ma}}} \bibnamefont{and}
  \bibinfo{author}{\bibfnamefont{J.~N.} \bibnamefont{{Fry}}},
  \bibinfo{journal}{\apj} \textbf{\bibinfo{volume}{543}}, \bibinfo{pages}{503}
  (\bibinfo{year}{2000}), \eprint{astro-ph/0003343}.

\bibitem[{\citenamefont{{Seljak}}(2000)}]{Seljak2Halo}
\bibinfo{author}{\bibfnamefont{U.}~\bibnamefont{{Seljak}}},
  \bibinfo{journal}{\mnras} \textbf{\bibinfo{volume}{318}},
  \bibinfo{pages}{203} (\bibinfo{year}{2000}), \eprint{astro-ph/0001493}.

\end{thebibliography}
